\documentclass[twocolumn,superscriptaddress,floatfix]{revtex4-2}
\usepackage{gnuplottex}
\usepackage{mathtools}
\usepackage{float}
\usepackage{array}
\usepackage[english]{babel}
\usepackage[utf8]{inputenc}
\usepackage{amsmath}
\usepackage[table]{xcolor}
\usepackage{amsfonts}
\usepackage{graphicx}
\usepackage{dsfont}
\usepackage[colorlinks=true, linkcolor=blue, citecolor=dred]{hyperref}
\usepackage{algorithm}
\usepackage[noend]{algpseudocode}
\usepackage{bbold}
\usepackage{mathtools}
\usepackage{amssymb}
\usepackage{bm}
\usepackage{bbm}
\usepackage{tabularx, booktabs}
\usepackage{xspace}
\usepackage{listings}
\usepackage{tikz}
\usepackage{bbding}
\usepackage{physics}
\usepackage{mciteplus}
\mciteErrorOnUnknownfalse
\usetikzlibrary{calc, positioning, quantikz}
\usepackage[noend]{algpseudocode}
\newcolumntype{Y}{>{\centering\arraybackslash}X}
\definecolor{dgreen}{rgb}{0,.5,0}
\definecolor{dblue}{rgb}{0,0,.5}
\definecolor{dred}{rgb}{0.5,0,.5}
\newcommand{\bfr}{\mathbf{r}}

\newcommand{\psh}[2]{\ensuremath{\langle #1|#2\rangle}\xspace}
\def\ddroit{{\rm d}}
\newcommand{\bmkappa}{{\bm \kappa}}
\newcommand{\bmtheta}{{\bm \theta}}

\newcommand{\bfH}{{\bf H}}

\newcommand{\bfG}{{\bf G}}

\newcommand{\rmA}{{\rm A}}
\newcommand{\rmB}{{\rm B}}
\DeclareMathOperator*{\argmin}{arg\,min}

\begin{document}

\title{Analytical non-adiabatic couplings and gradients within the state-averaged orbital-optimized variational quantum eigensolver}

\author{Saad Yalouz}
\email{yalouzsaad@gmail.com}
\affiliation{Laboratoire de Chimie Quantique, Institut de Chimie,
CNRS/Université de Strasbourg, 4 rue Blaise Pascal, 67000 Strasbourg, France}
\affiliation{Theoretical Chemistry, Vrije Universiteit, De Boelelaan 1083, NL-1081 HV, Amsterdam, The Netherlands}
\affiliation{Instituut-Lorentz, Universiteit Leiden, P.O. Box 9506, 2300 RA Leiden, The Netherlands}
\author{Emiel Koridon}
\email{emielkoridon@hotmail.com}
\affiliation{Theoretical Chemistry, Vrije Universiteit, De Boelelaan 1083, NL-1081 HV, Amsterdam, The Netherlands}
\affiliation{Instituut-Lorentz, Universiteit Leiden, P.O. Box 9506, 2300 RA Leiden, The Netherlands}
\author{Bruno Senjean}
\email{bruno.senjean@umontpellier.fr}
\affiliation{ICGM, Univ Montpellier, CNRS, ENSCM, Montpellier, France}
\author{Benjamin Lasorne}
\email{benjamin.lasorne@umontpellier.fr}
\affiliation{ICGM, Univ Montpellier, CNRS, ENSCM, Montpellier, France}
\author{Francesco Buda}
\affiliation{Leiden Institute of Chemistry,
Leiden University, Einsteinweg 55, P.O. Box 9502, 2300 RA Leiden, The Netherlands.}
\author{Lucas Visscher} 
\affiliation{Theoretical Chemistry, Vrije Universiteit, De Boelelaan 1083, NL-1081 HV, Amsterdam, The Netherlands}

\begin{abstract}
In this work, we introduce several technical and analytical extensions to our recent state-averaged orbital-optimized variational quantum eigensolver (SA-OO-VQE) algorithm (see Ref.~\citenum{Yalouz_2021}). 
Motivated by the limitations of current quantum computers,
the first extension consists in 
an efficient state-resolution procedure to find the SA-OO-VQE eigenstates, and not just the subspace spanned by them, while remaining in the equi-ensemble framework.
This approach avoids expensive intermediate resolutions of the eigenstates by postponing this problem to the very end of the full algorithm. 
The second extension allows for the estimation of analytical gradients and non-adiabatic couplings,
which are crucial in many practical situations ranging from the search of conical intersections to the simulation of quantum dynamics, in, for example, photoisomerization reactions.
The accuracy of our new implementations is demonstrated on the formaldimine molecule CH$_2$NH (a minimal Schiff base model relevant for the study of  photoisomerization in larger bio-molecules),
for which we also perform a geometry optimization to locate a conical intersection between the ground and first-excited electronic states
of the molecule.
\end{abstract}

\maketitle

\section{Introduction}

Many fundamental processes in nature, such as photosynthesis and vision, are triggered by light absorption. 
Thus, a proper description of the associated primary light-induced photochemical events requires a quantum-mechanical approach able to treat accurately both the ground and the excited electronic states. 
Although density functional theory (DFT) and its time-dependent extension to excited states (TDDFT) have seen huge progress in treating molecular and condensed matter systems near equilibrium~\cite{verma2020status,adamo2013calculations,doi:10.1021/acs.accounts.1c00312}, these approaches are not adequate to accurately describe photochemical reaction paths where the Born-Oppenheimer approximation breaks down for several strongly coupled electronic states that get very close in energy~\cite{rozzi2017quantum}. 
Especially couplings between the first excited and ground states are problematic because of the single-reference character of many quantum chemical methods (for instance the popular time-dependent density functional theory, TDDFT approach). 
While single-reference approaches with spin-flip excitations might help to overcome some of these limitations~\cite{minezawa2019trajectory}, in general more accurate and computationally demanding multi-configurational wavefunction approaches are required for modeling these intrinsically non-adiabatic cases. 
A good example  is the description of the prototypical photoisomerization process in the retinal chromophore of rhodopsin, one of the most studied events in photobiology~\cite{gozem2017theory}.
Schematically, after the initial photoexcitation, this event proceeds via the relaxation in the first excited state (S$_1$) towards a conical intersection (CoIn) region.
Here, the population is transferred back to the ground state (S$_0$) where the isomerization is completed. 
In order to describe dynamically this type of event, one needs the knowledge of the potential energy surfaces (PES) for the electronic states involved in the process, typically S$_0$ and S$_1$.
Moreover, one should also efficiently compute the gradient of the PES with respect to the nuclear displacement, which in a semiclassical non-adiabatic molecular dynamics scheme provides the forces driving the nuclear subsystem~\cite{rozzi2017quantum,gozem2017theory,agostini2016quantum}.
Finally, it is also crucial to estimate the non-adiabatic coupling terms between the two electronic states, which eventually determine the conical topography of the crossing between the two PES and the dynamical coupling that results in population transfer between the two states~\cite{faraji2018calculations,lee2021fast}.
The challenge in computational quantum chemistry is to obtain all these necessary ingredients at an affordable numerical cost and yet with good accuracy.

Methods that are able to provide both non-adiabatic couplings and a correct description of the PES topology and topography (double cone of dimension two with respect to variations of the nuclear coordinates) of conical intersections require, formally, that the problem be solved at the very end with a final Hamiltonian diagonalization.
When the crossing occurs between the first excited and ground states, this implies a democratic treatment of both wavefunctions within a common Slater determinant basis set, which in practice calls for a state-averaged (SA) orbital optimization.
This can be achieved in-principle by the state-averaged multiconfigurational self-consistent field (SA-MCSCF) method~\cite{helgaker2014molecular}.
In practice, the diagonalization step is the principal bottleneck and one has to consider small complete active spaces (CAS), thus leading to the state-averaged complete active space self-consistent field (SA-CASSCF) method~\cite{helgaker2014molecular}.
However, this decrease in complexity comes at the expense of a missing dynamical correlation treatment, that is
usually recovered by {multireference quasidegenerate perturbation techniques, as in the XMS-CASPT2~\cite{shiozaki2011communication}, XMCQDPT2~\cite{granovsky2011extended}, or QD-NEVPT2 methods~\cite{park2019analytical,angeli2004quasidegenerate}; see also Ref.~\citenum{gozem2014shape} for a comparative discussion of the correct treatment of degeneracies with a selection of excited-state approaches}.

With the advent of quantum computing,
the dream of a very large CAS becomes possible again,
thus turning small SA-CASSCF into large SA-CASSCF which should be good enough to account for a qualitatively correct description of the wave function and also include a substantial part of the (previously missing) so-called dynamical correlation.
Note that even with relatively small active spaces, the dynamical correlation can be retrieved a posteriori by other techniques on quantum computers, with no additional qubits or circuit depth, but at the expense of more measurements, as described by Takeshita {\it et al.}~\cite{takeshita2020increasing}.
Recently, the \textit{quantum analogue} of SA-CASSCF
has been introduced by Yalouz \textit{et al.}~\cite{Yalouz_2021} based on a state-averaged orbital-optimized (SA-OO) extension of the
variational quantum eigensolver (VQE) algorithm~\cite{peruzzo2014variational,mcclean2016theory}, thus referred to as the SA-OO-VQE algorithm.
While SA-OO-VQE has been shown to provide
an accurate and democratic description of both the ground and first-excited PES~\cite{Yalouz_2021},
its extension to excited-state quantum dynamics
requires the knowledge of energy gradients and non-adiabatic couplings.
In this work, we show how these properties can be
analytically estimated on a quantum computer
within the SA-OO-VQE framework, following the coupled-perturbed equations~\cite{lengsfield1984evaluation, staalring2001analytical,lengsfield1992nonadiabatic,yarkony1995modern,snyder2015atomic,snyder2017direct, fdez2016analytical}.
In analogy with Ref.~\citenum{Yalouz_2021},
the performance of our algorithm is illustrated
on the minimal Schiff base model (\textit{i.e.} the formaldimine molecule), for which results are indistinguishable from its classical analogue, the (coupled-perturbed) SA-CASSCF method.

The paper is organized as follows.
For pedagogical purposes, we briefly introduce quantum chemistry for excited states in Sec.~\ref{subsec:QC_excited},
from the Born--Openheimer approximation
in Sec.~\ref{subsubsec:BO} to the SA-MCSCF method in Sec.~\ref{subsubsec:SAMCSCF}.
Turning to quantum computing
in Sec.~\ref{subsec:gradients_NAC_QC},
a summary of the SA-OO-VQE in given in Sec.~\ref{subsubsec:SA-OO-VQE}, and
a way to extract the eigenstates (\textit{i.e.} the adiabatic states) is provided in Sec.~\ref{subsubsec:control}, where we also discuss the alternative choice of having diabatic or adiabatic states within the SA-OO-VQE algorithm.
The analytical estimation of energy gradients and non-adiabatic couplings is then described in Secs.~\ref{subsubsec:analytical_gradients}
and \ref{subsubsec:NAC}, respectively, and they are compared
with classical methods in Sec.~\ref{subsec:gradients_NAC_results}.
Using the equations for the analytical gradients, a geometry optimization to the degeneracy point
is executed in Sec.~\ref{subsec:geom_opt} as a simple illustration.
A more involved optimization to the minimal energy crossing point (MECI) that requires the knowledge of non-adiabatic couplings is performed in Sec.~\ref{sec:MECI_opt}.
Conclusions and perspectives are finally discussed in
Sec.~\ref{sec:conclu}.

\section{Theory}\label{sec:theory}

\subsection{Quantum Chemistry for excited states}\label{subsec:QC_excited}

\subsubsection{Born--Oppenheimer and the adiabatic approximation}\label{subsubsec:BO}

One of the most fundamental approximations used in theoretical chemistry is the adiabatic approximation between electrons and nuclei, which most often takes the form of the Born--Oppenheimer approximation and sometimes of the Born--Huang approximation (the latter being essentially used for highly-accurate treatments of vibrations in small molecules). 
In both cases, non-adiabatic couplings due to the action of the kinetic energy operator of the nuclei on the parametric dependence of the adiabatic electronic wavefunctions are neglected; 
however, the Born--Huang approximation considers nuclear-mass-dependent diagonal corrections that are to be added to the potential energy surface obtained as a single adiabatic eigenvalue of the clamped-nucleus Hamiltonian.

Such approximations are justified by the small ratio of electronic over nuclear masses, which results in very different energy and time scales in the vast majority of cases. 
However, electronic degeneracies may occur at certain nuclear geometries (Jahn--Teller crossings due to symmetry, or, more generally, conical intersections). 
At such points, the two intersecting potential energy surfaces take locally the shape of a double cone (over a subspace of dimension 2 for a two-state crossing). 
The two nuclear displacements that lift degeneracy to first order are usually called branching-space vectors; their directions can be identified to the energy gradient difference and first-order non-adiabatic coupling (NAC) vector. 
Formally, the $x$-component of the NAC vector between two electronic states $\ket{\Psi_I}$ and $\ket{\Psi_J}$ is defined by
\begin{equation}\label{eq:def_NAC}
D_{IJ} = \bra{\Psi_I} \ket{\pdv{}{x} \Psi_J}.
\end{equation}
where $x$ represents a given nuclear coordinate and the wavefunctions depend parametrically on it (integration, however, is performed over the electronic Hilbert space only). If the wavefunctions considered in Eq.~(\ref{eq:def_NAC}) are exact, they yield
\begin{equation}\label{eq:form_NAC_HF}
D_{IJ} = \frac{1}{ E_J - E_I } \bra{\Psi_I} \pdv{\hat{\mathcal{H}}}{x} \ket{\Psi_J},
\end{equation}
in virtue of the  off-diagonal Hellmann--Feynman theorem.
The magnitude of the NAC vector is ill-defined at a conical intersection, since it diverges as the inverse of the energy difference [see Eq.~(\ref{eq:form_NAC_HF})].
The numerator, however, is well-defined and often called the derivative coupling vector; note that the nomenclature is not fixed in the literature. It can be viewed as a transition gradient. The other vector that forms the branching space together with the derivative coupling is the gradient (half) difference,
\begin{equation}\label{eq:def_GD}
G_{IJ} = \frac{1}{2} \left( \bra{\Psi_J} \pdv{\hat{\mathcal{H}}}{x} \ket{\Psi_J} - \bra{\Psi_I} \pdv{\hat{\mathcal{H}}}{x} \ket{\Psi_I} \right).
\end{equation}
The vectors $G_{IJ}$ and $(E_J - E_I)D_{IJ}$ -- often denoted $g$ and $h$ vectors or $x_1$ and $x_2$ vectors in this context -- play symmetrical roles: they form the two directions that make the adiabatic energy difference increase to first order from zero at a conical intersection. They actually are undetermined up to within a mutual rotation, which directly reflects the freedom in defining two specific degenerate eigenstates (see, e.g., Ref.~\citenum{gatti2017vibronic}).

In addition to being essential for the correct capture of the conical topography of crossings, non-adiabatic couplings are required for describing the coupled equations that govern the nuclear components of the molecular wavefunction. 
As already pointed out, they become large when the energy gap between electronic states decreases, which is why conical intersections are key for describing radiationless processes whereby population is transferred among electronic states. 
In practice, non-adiabatic quantum dynamics is often better described in terms of quasidiabatic electronic states that result from a unitary transformation of a relevant subset of coupled adiabatic states. 
They vary smoothly enough with respect to nuclear coordinates to allow for neglect of kinetic couplings but introduce instead nonzero potential couplings.

Further on this is beyond the scope of the present work and the literature on the subject is vast. 
We refer for example to Ref.~\citenum{baer2006beyond} for a comprehensive review of relevant concepts. 
Finally, let us stress that the ability of a computational method to describe correctly the topography of a conical intersection is intimately related to the formal possibility of using analytic derivative techniques for evaluating non-adiabatic couplings~\cite{gozem2014shape}. 
This somewhat relies on the fact that the final step of the whole computational procedure should be viewed as a Hamiltonian submatrix diagonalization that provides several eigenstates democratically within the same subspace. 
The state-averaged multi-configurational self-consistent-field (SA-MCSCF) method is an evident option in this context, with analytic derivatives applied similarly to diagonal and off-diagonal terms~\cite{lengsfield1992nonadiabatic}.

\subsubsection{State-averaged multi-configurational self-consistent-field method (SA-MCSCF)}\label{subsubsec:SAMCSCF}

The electronic structure Hamiltonian reads, in second quantization,
\begin{eqnarray}
\hat{\mathcal{H}}
=
\sum_{pq} h_{pq} \hat{E}_{pq} + \dfrac{1}{2}
\sum_{pqrs}
g_{pqrs}\hat{e}_{pqrs},
\end{eqnarray}
where the one- and two-electron integrals are defined as (in real algebra)
\begin{eqnarray}
h_{pq} = \int \, \ddroit \bfr \phi_p(\bfr) \left( - \dfrac{1}{2}\grad^2_\bfr + v_{ne}(\bfr) \right)
\phi_q(\bfr),
\end{eqnarray}
and
\begin{eqnarray}
g_{pqrs} = \iint \ddroit \bfr_1 \ddroit \bfr_2\, \dfrac{\phi_p(\bfr_1)\phi_r(\bfr_2)\phi_q(\bfr_1)\phi_s(\bfr_2)}{|\bfr_1 - \bfr_2|},
\end{eqnarray}
respectively,
where $\lbrace \phi_p \rbrace$ are the molecular orbitals defining the (finite) basis set and $v_{ne}(\bfr)$ is the nucleus-electron potential.
The one- and two-body spin-free excitation operators are defined such as $\hat{E}_{pq} = \sum_{\sigma} \hat{a}_{p\sigma}^\dagger \hat{a}_{q\sigma}$
and $ \hat{e}_{pqrs} = \sum_{\sigma,\tau} \hat{a}_{p\sigma}^\dagger \hat{a}_{r\tau}^\dagger \hat{a}_{s\tau}\hat{a}_{q\sigma}$
where $\hat{a}_{p\sigma}^\dagger$ ($\hat{a}_{p\sigma}$) is the creation (annihilation) operator of an electron with spin $\sigma$ in spatial orbital $p$.
Due to the exponential increase of the configuration space in the number of molecular orbitals,
it is of common use to select only a restricted (and, ideally, relevant) part of it in practical calculation, for instance by considering the active space approximation where
the orbital space is separated into a set of frozen occupied, active and virtual orbitals.
In such a reduced configuration space,
the configuration interaction method is not invariant anymore under orbital rotations~\cite{siegbahn1981complete,helgaker2014molecular} and the choice of orbitals will influence the quality of the result.
Hence, one has to consider the re-optimization of the orbitals, thus leading to the MCSCF model which wavefunction reads:
\begin{eqnarray}\label{eq:MCSCF_wavefunction}
\ket{\Psi(\bmkappa,\mathbf{c})} = e^{- \hat{\kappa}} \left(\sum_i c_{i} \ket{\Phi_i}\right),
\end{eqnarray}
where $\lbrace \ket{\Phi_i} \rbrace$ are Slater determinants or configuration state functions,
and $\hat{U}_\text{O}(\bmkappa) = e^{-\hat{\kappa}}$ is the orbital-rotation operator.
The latter is defined as follows in the spin-restricted formalism with real algebra:
\begin{eqnarray}\label{eq:orb_rot}
\hat{\kappa} = \sum_{p>q}^\text{MOs} \kappa_{pq}(\hat{E}_{pq}-\hat{E}_{qp}).
\end{eqnarray}
The parameters of the wavefunction in Eq.~(\ref{eq:MCSCF_wavefunction})
are determined by variationally optimizing the expectation value of the energy:
\begin{eqnarray}
E = \min_{\bmkappa,\mathbf{c}} \dfrac{\bra{\Psi(\bmkappa,\mathbf{c})}\hat{\mathcal{H}}\ket{\Psi(\bmkappa,\mathbf{c})}}{\psh{\Psi(\bmkappa,\mathbf{c})}{\Psi(\bmkappa,\mathbf{c})}}.
\end{eqnarray}
In order to have a democratic description of ground and excited states, one can simultaneously optimize several MCSCF states that are generated from the same orbital basis.
As extensively discussed in Ref.~\citenum{helgaker2014molecular},
it is convenient to introduce an exponential unitary parametrization of the configuration space with nonredundant variables,
\begin{eqnarray}
\hat{U}_\text{C}(\mathbf{S}) = e^{-\hat{S}},
\end{eqnarray}
where
\begin{eqnarray}\label{eq:S}
\hat{S} = \sum_{J} \sum_{K>J} S_{KJ} \left( \ket{\Psi_K^{(0)}}\bra{\Psi_J^{(0)}} - \ket{\Psi_J^{(0)}}\bra{\Psi_K^{(0)}}\right)
\end{eqnarray}
and
\begin{eqnarray}
\ket{\Psi_I^{(0)}} = \sum_i c_{Ii}^{(0)}\ket{\Phi_i}
\end{eqnarray}
are initial orthonormal states built from the same set of molecular orbitals.
Within the SA-MCSCF model, the wavefunctions are
subject to a double-exponential parametrization
\begin{eqnarray}\label{eq:SAMCSCF_wavefunction}
\ket{\Psi_I(\bmkappa,\mathbf{S})} = e^{- \hat{\kappa}} e^{-\hat{S}} \ket{\Psi_I^{(0)}},
\end{eqnarray}
where, according to the generalization of the Rayleigh--Ritz variational principle for an ensemble of ground and excited states~\cite{gross1988rayleigh}, the parameters are variationally optimized by minimizing the state-averaged energy 
\begin{eqnarray}\label{eq:SA-MCSCF_energy}
E^{\text{SA-MCSCF}} = \min_{\bmkappa,\mathbf{S}} \sum_I w_I\bra{\Psi_I(\bmkappa,\mathbf{S})}\hat{\mathcal{H}}\ket{\Psi_I(\bmkappa,\mathbf{S})},
\end{eqnarray}
where 
$\sum_I w_I = 1$
and the states are automatically orthonormalized
as they are generated from unitary transformations of the initial orthonormal states $\lbrace \ket{\Psi_I^{(0)}}\rbrace$.
Note that due to the orbital optimization, the converged individual and state-averaged energies may vary with the weights.
In practice, the equal weight SA-MCSCF (where all weights are equal) is usually considered.
Finally, the dependence on $\bmkappa$ in the wavefunctions
can actually be transferred to the electronic integrals in the Hamiltonian, \textit{i.e.}
$h_{pq} \rightarrow h_{pq}(\bmkappa)$
and $g_{pqrs} \rightarrow g_{pqrs}(\bmkappa)$,
such that Eq.~(\ref{eq:SA-MCSCF_energy}) equivalently reads
\begin{eqnarray}\label{eq:SA-MCSCF_energy_2}
E^{\text{SA-MCSCF}} = \min_{\bmkappa,\mathbf{S}} \sum_I w_I\bra{\Psi_I(\mathbf{S})}\hat{\mathcal{H}}(\bmkappa)\ket{\Psi_I(\mathbf{S})},
\end{eqnarray}
where
$\hat{\mathcal{H}}(\bmkappa) =\hat{U}_\text{O}^\dagger (\bmkappa)\hat{\mathcal{H}}\hat{U}_\text{O}(\bmkappa)$ is the MO-basis transformed Hamiltonian.

While the SA-MCSCF method allows for a democratic description of ground and excited states,
it is only variational with respect to the state-averaged energy, so that an individual state is not variational.
This makes the calculation of analytical energy gradients of each individual state more complicated,
as it requires the introduction of specific Lagrangians and the solution of so-called coupled-perturbed equations,
as further discussed in Sec.~\ref{subsubsec:analytical_gradients}.

\subsection{Estimation of energies, analytical gradients and non-adiabatic coupling on a quantum computer}\label{subsec:gradients_NAC_QC}

\subsubsection{State-averaged orbital-optimized variational-quantum-eigensolver (SA-OO-VQE)}\label{subsubsec:SA-OO-VQE}

The variational quantum eigensolver (VQE)~\cite{peruzzo2014variational,mcclean2016theory} represents one of the most promising methods to estimate the ground-state energy on near-term quantum computers.
As suggested by the name of the algorithm, the VQE relies on the Rayleigh--Ritz variational principle and
consists in finding the closest approximation to the ground-state wavefunction
thanks to a given \textit{ansatz} (defined by a parametrized unitary operation $\hat{U}(\bmtheta)$).
Applying this unitary operation to a chosen initial state (usually very easy to prepare, such as the Hartree--Fock (HF) Slater determinant $\ket{\text{HF}}$) leads to a parametrized trial wavefunction $\ket{\Psi(\boldsymbol{\theta})} = \hat{U}(\bmtheta)\ket{\text{HF}}$, from which the associated energy is estimated by repeated measurements of the quantum circuit.
Unfortunately, the extension of the VQE algorithm to excited state is not trivial, as a variational estimation of the excited-state energies can only be defined under orthogonal constraints.
Such constraints have been considered by adding penalization terms to the Hamiltonian, thus leading to the state-specific variational quantum deflation (VQD) algorithm~\cite{higgott2019variational,jones2019variational,jouzdani2019method,ibe2020calculating}
where each state is determined by a separate minimization (or only two minimizations in total if the first one is performed on a state-average ensemble~\cite{wen2021variational}).
Other extensions can treat excited states on the same footing, but still favor the ground state~\cite{mcclean2017hybrid,colless2018computation,ollitrault2019quantum,motta2020determining}.
However, the proper description of conical intersections or avoided crossings require a democratic description of both the ground and excited states.
Such an equal footing treatment can be achieved by performing a single minimization (or resolution) for all states sharing the same ansatz, as in multistate-contracted VQE (MC-VQE)~\cite{parrish2019quantum,parrish2019hybrid}, fully-weighted subspace-search VQE (SS-VQE)~\cite{nakanishi2019subspace}, variance-VQE~\cite{zhang2020variational}
and the quantum filter diagonalization method~\cite{parrish2019quantumfilter,bespalova2020hamiltonian}.
Inspired by the SS-VQE method of Nakanishi \textit{et al.}~\cite{nakanishi2019subspace}, 
we proposed the (equi-weighted)
state-averaged orbital-optimized VQE (SA-OO-VQE), that can be seen as a combination of
a state-averaged VQE (SA-VQE) and a state-averaged orbital-optimization (SA-OO) procedure.
Let us briefly summarize each step of the SA-OO-VQE, focusing on an equi-ensemble of two-states (the extension to more electronic states is straightforward).
\begin{enumerate}
\item \textbf{{Initialization:} } Initialize the circuits with two orthonormal states $\ket{\Phi_{\rm A}}$ and $\ket{\Phi_{\rm B}}$.\\
\item \textbf{{SA-VQE:}}  Apply a quantum ansatz (\textit{i.e.} a given quantum circuit) to transform both initial states into trial states  $\ket{\Psi_{\rm A}(\boldsymbol{\theta})} = \hat{U}(\boldsymbol{\theta})|\Phi_{\rm A}\rangle$ and $\ket{\Psi_{\rm B}(\boldsymbol{\theta})}= \hat{U}(\boldsymbol{\theta})|\Phi_{\rm B}\rangle$, and find
the optimal set of ans\"atze parameters that minimizes the state-averaged energy 
\begin{eqnarray}
\boldsymbol{\theta}^* = \argmin_{\boldsymbol{\theta}} E^\text{SA-OO-VQE}(\bmkappa,\boldsymbol{\theta})
\end{eqnarray}
for a fixed orbital basis $\bmkappa$, where the state-average energy reads
\begin{eqnarray}\label{eq:SAOOVQE_energy}
E^{\text{SA-OO-VQE}}(\bmkappa,\bmtheta) = w_{\rm A} \langle\Psi_{\rm A} (\boldsymbol{\theta})|\hat{\mathcal{H}} (\boldsymbol{\kappa})|\Psi_{ \rm A} (\boldsymbol{\theta})\rangle
\nonumber \\
+ w_{\rm B} \langle\Psi_{\rm B} (\boldsymbol{\theta})|\hat{\mathcal{H}} (\boldsymbol{\kappa})|\Psi_{ \rm B} (\boldsymbol{\theta})\rangle
\end{eqnarray}
with $w_{\rm A}$ and $w_{\rm B}$ the weights attributed to each state with the normalization condition $w_{\rm A}+w_{\rm B}=1$.
Note that this energy is lower-bounded by the ensemble energy of the exact two lowest eigenstates (denoted by $\ket{\Psi_0}$ and $\ket{\Psi_1}$) of  $\hat{\mathcal{H}}(\bmkappa^*)$ in the active-space approximation, according the variational principle~\cite{gross1988rayleigh}.
\item \textbf{{SA-OO:}} Rotate the orbital basis to find the optimal set of parameters that minimize the state-averaged energy 
\begin{eqnarray}
\bmkappa^* = \argmin_{\bmkappa} E^\text{SA-OO-VQE}(\bmkappa,\boldsymbol{\theta})
\end{eqnarray}
(\textit{e.g.} with Newton-Raphson), for a fixed set of parameters $\boldsymbol{\theta}$.\\
\item \textbf{{SA-OO-VQE:}} Repeat steps 2 and 3 until the state-average energy is minimized with respect to both $\boldsymbol{\theta}$ and $\bmkappa$, \textit{i.e.} find 
\begin{eqnarray}
(\bmkappa^*,\boldsymbol{\theta}^*) = \argmin_{\bmkappa,\boldsymbol{\theta}} E^\text{SA-OO-VQE}(\bmkappa,\boldsymbol{\theta}).
\end{eqnarray}
\end{enumerate}

As discussed in previous works~\cite{nakanishi2019subspace,Yalouz_2021}, the
lower bound in Eq.~(\ref{eq:SAOOVQE_energy}) is uniquely defined if $w_A > w_B$, but is invariant under any rotation between $\ket{\Psi_0}$
and $\ket{\Psi_1}$ in the equi-ensemble case ($w_A = w_B$).
Hence, considering the case $w_A = w_B$ does not guarantee that the optimized states $\ket{\Psi_A(\bmtheta^*)} $ and $\ket{\Psi_B(\bmtheta^*)}$ are the closest approximation of the eigenstates $\ket{\Psi_0}$ and $\ket{\Psi_1}$.
However, this enforces the definition of a well-defined two-state subspace spanned by either $\Psi_A$ and $\Psi_B$ or $\Psi_0$ and $\Psi_1$, such that the latter are eigenstates.
Forcing this correspondence (that we refer to as the \textit{state-resolution})
is a complicated task that can be handled in different ways.
Considering $w_A > w_B$ is a straightforward solution, but
this constraint may complicate the SA-VQE optimization considerably~\cite{nakanishi2019subspace}.
Additional tricks can be used in the equi-ensemble case,
by considering additional cost-functions to be maximized~\cite{nakanishi2019subspace},
a classical diagonalization~\cite{parrish2019quantum},
or another type of cost-functions that use the variance of the states~\cite{zhang2020variational}. 
In Sec.~\ref{subsubsec:control}, we discuss another approach, inspired by the one of Nakanishi~\cite{nakanishi2019subspace}, to solve the state-resolution of a two-state ensemble within the SA-OO-VQE algorithm, when the initial states are the HF Slater determinant and any singlet singly-excited configuration interaction (CIS) state.
Note that while we focus on those particular initial states in this manuscript, any other choice could in principle be considered.

\subsubsection{State-resolution procedure}\label{subsubsec:control}

In this section,
we propose another method to capture the active-space eigenvectors of $\hat{\mathcal{H}}(\bmkappa)$), which requires few additional gates
and a negligible increase in the number of measurements.
\begin{figure}
    \centering
    \includegraphics[width=0.9\columnwidth]{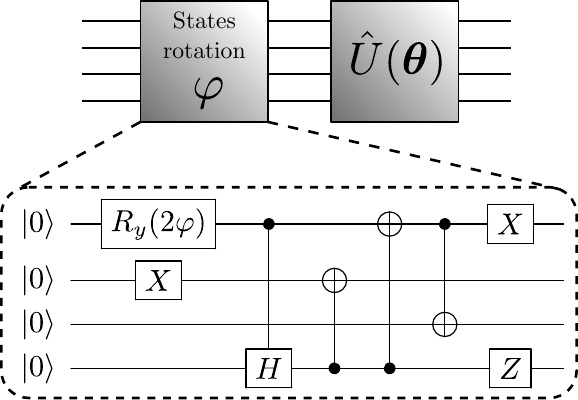}
    \includegraphics[width=\columnwidth]{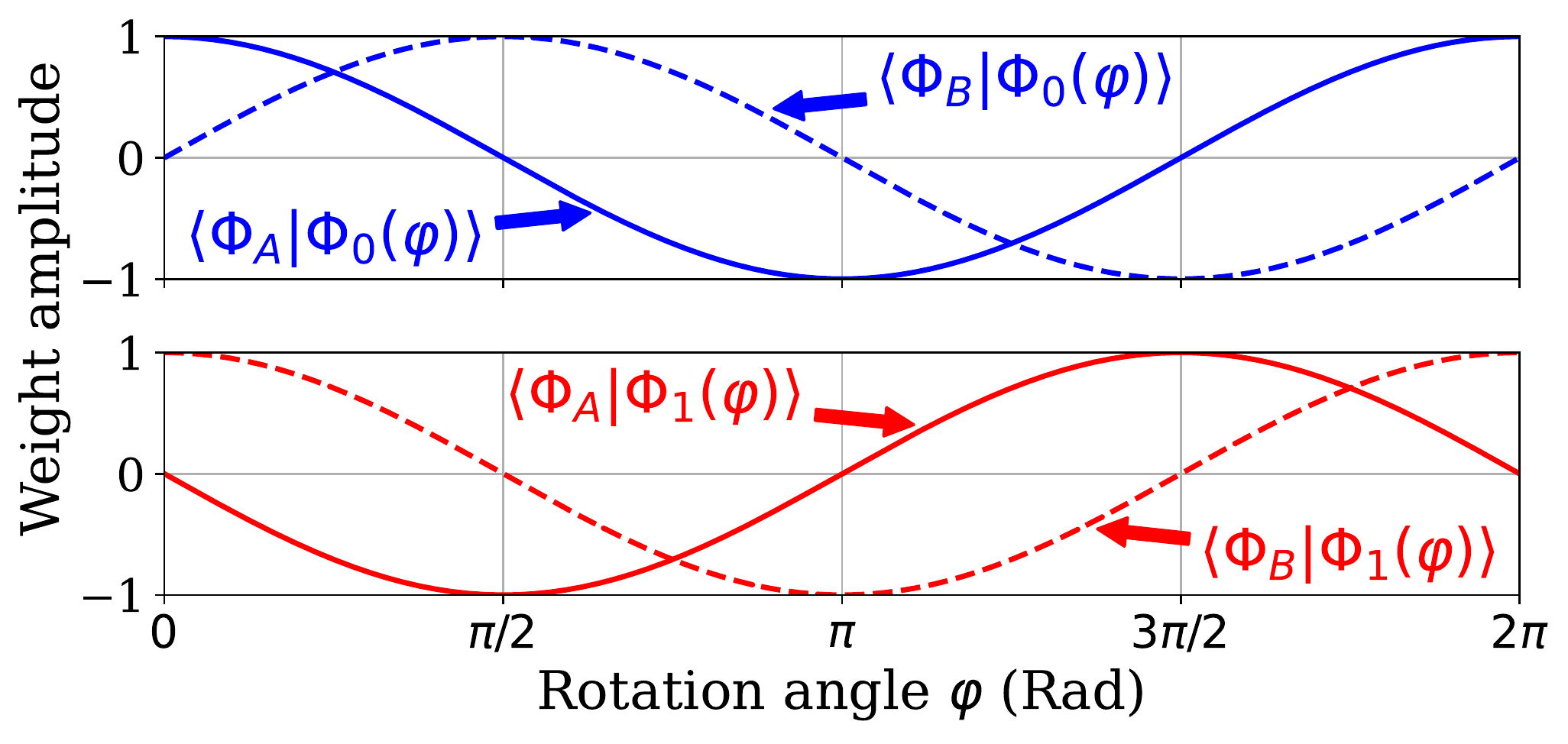}
    \caption{\textbf{Upper panel:} Short-depth quantum circuit specifically designed for the state-resolution of the SA-OO-VQE algorithm, to build the state $\ket{\Phi(\varphi)} = \cos{\varphi} \ket{\Phi_{\rm A}} + \sin{\varphi} \ket{\Phi_{\rm B}}$,
    i.e. a rotation between the HF state
    $ \ket{\Phi_{\rm A}} = | \text{HF} \rangle$
    and the HOMO-LUMO singlet CIS state $ \ket{\Phi_{\rm B}} = \hat{E}_\text{hl}| \text{HF} \rangle /\sqrt{2}$.
    \textbf{Lower panel:} Overlaps $ \langle    \Phi_A | \Phi_{0/1}(\varphi) \rangle$ and $ \langle  \Phi_B | \Phi_{0/1}(\varphi) \rangle $ as a function of the rotation parameter $\varphi$ (with $ \ket{\Phi_{\rm A}} = | \text{HF} \rangle$
    and  $ \ket{\Phi_{\rm B}} = \hat{E}_\text{hl}| \text{HF} \rangle /\sqrt{2}$).}
    \label{fig:resolve}
\end{figure}
Our approach consists in taking advantage of the rotational invariance of the equi-ensemble state-averaged energy, in order to postpone the state-resolution to the very end of the SA-OO-VQE algorithm. 
Considering the equi-ensemble (\textit{i.e.} $w_\text{A} = w_\text{B} $), after convergence of the SA-OO-VQE algorithm, the resulting Hilbert space spanned by $\ket{\Psi_{\rm A}(\bmtheta^*)}$ and $\ket{\Psi_{\rm B}(\bmtheta^*)}$ is a good approximation to the 
subspace spanned by the SA-CASSCF states -- classical analogue of the SA-OO-VQE method --, within the same active space.
(As discussed in Sec.~\ref{subsubsec:SA-OO-VQE}, the SA-OO-VQE states are not constrained to be the eigenvectors of $\hat{\mathcal{H}}$, or, equivalently, to form the adiabatic basis that diagonalizes $\hat{\mathcal{H}}(\bmkappa^*)$).
To resolve the ground and first excited SA-OO-VQE \textit{eigenstates} (which should be good approximations to $\ket{\Psi_{\rm 0}^\text{SA-CASSCF}}$ and $\ket{\Psi_{\rm 1}^\text{SA-CASSCF}}$),
we propose to implement a rotation between the initial states $\ket{\Phi_{\rm A}}$ and $\ket{\Phi_{\rm B}}$, such that the new rotated initial states become
\begin{equation}\label{eq:rotated_state_phi}
\begin{split}
    &| \Phi_0(\varphi) \rangle = \cos{\varphi} |\Phi_{ \rm A}\rangle + \sin{\varphi} |\Phi_{ \rm B} \rangle,  \\
    &| \Phi_1(\varphi) \rangle  = -\sin{\varphi} |\Phi_{ \rm A}\rangle + \cos{\varphi} |\Phi_{ \rm B}\rangle,
\end{split}
\end{equation}
where we have the property $| \Phi_1(\varphi) \rangle = | \Phi_0(\varphi+\pi/2) \rangle$. 
After applying the ansatz with optimized parameters $\boldsymbol{\theta^*}$, these new rotated initial states evolve to
\begin{equation}\label{eq:rotated_state}
\begin{split}
    &| \Psi_0(\varphi,\boldsymbol{\theta^*}) \rangle = \cos{\varphi} |\Psi_{ \rm A} (\boldsymbol{\theta^*})\rangle + \sin{\varphi} |\Psi_{ \rm B} (\boldsymbol{\theta^*})\rangle,\\  
    &| \Psi_1(\varphi,\boldsymbol{\theta^*}) \rangle = -\sin{\varphi} |\Psi_{ \rm A} (\boldsymbol{\theta^*})\rangle + \cos{\varphi} |\Psi_{ \rm B} (\boldsymbol{\theta^*})\rangle, 
\end{split}
\end{equation}
which leads to a rotation between the final SA-OO-VQE states.
Note that they remain orthonormal and, by virtue of the equi-ensembles properties, lead to the same state-averaged energy that is invariant with respect to $\varphi$.
The state-resolution amounts to finding the value $\varphi \rightarrow \varphi^*$ that minimizes the energy of $| \Psi_0(\varphi,\boldsymbol{\theta^*}) \rangle$,
\begin{eqnarray}\label{eq:varphi}
\varphi^* = \argmin_\varphi \bra{\Psi_0(\varphi,\boldsymbol{\theta}^*)}\hat{\mathcal{H}}(\bmkappa^*)\ket{\Psi_0(\varphi,\boldsymbol{\theta}^*)}
\end{eqnarray}
(or, equivalently, maximizes the one of 
$| \Psi_1(\varphi,\boldsymbol{\theta^*}) \rangle$),
thus making both $| \Psi_0(\varphi^*,\boldsymbol{\theta^*}) \rangle$ and $| \Psi_1(\varphi^*,\boldsymbol{\theta^*}) \rangle$ approximated eigenstates of $\hat{\mathcal{H}}(\bmkappa^*)$.
Satisfying Eq.~(\ref{eq:varphi}) can be seen as the fifth step of the SA-OO-VQE algorithm (see Sec.~\ref{subsubsec:SA-OO-VQE} for the first four steps).
In Fig.~\ref{fig:resolve}, we show the short-depth circuit we specifically developed to perform the
rotation between the HF determinant $\ket{\Phi_\rmA} = \ket{\text{HF}}$ and a singlet-excited CIS state $\ket{\Phi_\rmB} = - \hat{E}_ {\rm hl}\ket{\text{HF}}/\sqrt{2}$, where `h' and `l' refer to the HOMO and LUMO orbitals, respectively.
In practice, the circuit works as follows
(for the sake of simplicity, we focus on a 2-spatial-orbital (4-spin-orbital) -- \textit{i.e.} 4-qubit -- and 2-electron CAS).
Starting with the 4 qubits in the $\ket{0}$ state,
a $R_y(2\varphi)$ rotation gate and a $X$ gate
are applied to the first and second qubit, respectively,
thus leading to the quantum superposition 
\begin{eqnarray}\label{eq:RyX}
R_y^0(2\varphi)X^1 \ket{0000} =  \cos{\varphi}\ket{0100} + \sin{\varphi}\ket{1100}.
\end{eqnarray}
Then, a controlled-Hadamard gate 
transforms the state $\ket{1100}$ into
$\frac{1}{\sqrt{2}}(\ket{1100} + \ket{1101})$,
which evolves into 
$\frac{1}{\sqrt{2}}(\ket{1110} + \ket{0001})$
after applying three CNOT gates.
The first term in the right hand side of Eq.~(\ref{eq:RyX}) remains invariant with respect to the aforementioned operations, such that the state now reads
\begin{eqnarray}
\cos{\varphi}\ket{0100} + \frac{\sin{\varphi}}{\sqrt{2}}(\ket{1110} + \ket{0001}).
\end{eqnarray}
Finally, we apply $X$ and $Z$ gates to the first and last qubits, respectively, to arrive at the final expression
\begin{eqnarray}\label{eq:rotation_circuit}
\ket{\Phi_0(\varphi)}&=&\cos{\varphi} \ket{1100} +  \frac{\sin{\varphi}}{\sqrt{2}}\left( \ket{0110} - \ket{1001}\right)\nonumber \\
&=& \cos{\varphi} \ket{\Phi_\rmA} + \sin{\varphi}\ket{\Phi_\rmB},
\end{eqnarray}
where $| \Phi_{\rm A} \rangle = | \text{HF}\rangle$
and $| \Phi_{\rm B} \rangle = - \hat{E}_\text{hl}| \text{HF} \rangle /\sqrt{2}  = (\ket{0110}-\ket{1001})/\sqrt{2}$ is the HOMO--LUMO singlet-excited CIS state.
Replacing $\varphi \rightarrow \varphi + \pi/2$ in Eq.~(\ref{eq:rotation_circuit}), one recovers $\ket{\Phi_1(\varphi)}$ in
Eq.~(\ref{eq:rotated_state_phi}),
such that the parameter $\varphi$ can be tuned to realize any  real linear combination between $| \Phi_{\rm A} \rangle$ and $| \Phi_{\rm B} \rangle$, as
illustrated in the lower panels of Fig.~\ref{fig:resolve}.  
Note that this circuit is valid for any singlet-excited CIS state 
$\ket{\Phi_\rmB}$, by simply applying the quantum gates to the qubits associated to the orbitals involved in the excitation.

Note that the idea introduced here for the state-resolution procedure in SA-OO-VQE follows closely the one proposed by Nakanishi \textit{et al.} (Sec II.A. of Ref.~\citenum{nakanishi2019subspace}).
Indeed, the additional circuit in Fig.~\ref{fig:resolve} is equivalent to their additional unitary operation $V(\phi)$, for which we provide an explicit form
for any two-state ensemble (with a specific focus on initial states that are the HF and any singlet-excited CIS states).
Note also that a SA-OO and SA-VQE algorithms are alternatively employed in our method. 
The resulting SA-OO-VQE subspace is then more meaningful in terms of electronic correlations as it (ideally) provides analog results as in the SA-CASSCF method, contrary to the SS-VQE scheme that is equivalent to the CASCI method.
Because we work with an equi-ensemble, the state-resolution can be performed at the very end of the SA-OO-VQE algorithm only.
This attractive feature of the equi-ensemble SA-OO-VQE allows in principle to spare a lot of unnecessary quantum resources, as one can still end up with the (approximate) eigenstates
without requiring harder optimization procedures or additional quantum measurements at each instance of the SA-VQE algorithm.

\subsubsection{Analytical gradients}\label{subsubsec:analytical_gradients}

{Molecular properties can be accessed by estimating energy gradients with respect to a given perturbation~\cite{jensen2017introduction}. 
Analytical expressions, that are cheaper and more precise than finite difference techniques, have been derived in the context of ground-state VQE in Refs.~\cite{mitarai2020theory,obrien2019calculating,azad2021quantum,sokolov2021microcanonical}.}
In this section, we turn towards the question of the analytical evaluation of individual-state nuclear energy gradient with the SA-OO-VQE algorithm (which will be noted $\ket{\Psi_I}$, with $I=0,1,\ldots$). 
As opposed to the state-specific orbital-optimized VQE
introduced by Mizukami \textit{et al.}~\cite{mizukami2019orbital} and Takeshita {\it et al.}~\cite{takeshita2020increasing},
each set of variational parameters (in our case $\boldsymbol{\theta}$ and $\boldsymbol{\kappa}$) is not optimized to minimize each individual-state energy,
\begin{eqnarray}\label{eq:E_I_notvariational}
\dfrac{\partial E_I(\bmkappa,\bmtheta)}{\partial \kappa_{pq}} \neq 0,
\dfrac{\partial E_I(\bmkappa,\bmtheta)}{\partial \theta_{n}} \neq 0,
\end{eqnarray}
but rather to minimize the state-averaged energy,
\begin{eqnarray}\label{eq:SA_variational}
\dfrac{\partial E_\text{SA}(\bmkappa,\bmtheta)}{\partial \kappa_{pq}} = 
\dfrac{\partial E_\text{SA}(\bmkappa,\bmtheta)}{\partial \theta_n} = 0,
\end{eqnarray}
{where in both Eq.~\eqref{eq:E_I_notvariational} and Eq.~\eqref{eq:SA_variational} it is implicit that the gradients are evaluated at the converged parameters.} This renders the estimation of the individual-state nuclear energy gradients more complicated, as it has to take into account the non-variational character of the method.
Fortunately, one can build analytical Lagrangians that are fully variational with respect to every parameter~\cite{helgaker1984second}, such that their optimization
facilitates the estimation of the targeted quantities (\textit{e.g.} energy derivatives and non-adiabatic couplings in our case).

Following this strategy, we build an individual-state Lagrangian $\mathcal{L}_I$ that depends on all the parameters as follows,
\begin{eqnarray}\label{eq:CP_SAOOVQE_energy_grad}
\mathcal{L}_I
= E_I  + \left( \sum_{pq} \overline{\kappa}_{pq}^I \dfrac{\partial E_\text{SA}}{\partial \kappa_{pq}} - 0 \right)
+ \left( \sum_n \overline{\theta}_n^I \dfrac{\partial E_\text{SA}}{\partial \theta_n} - 0 \right).\nonumber\\
\end{eqnarray}
Note that, based on the state-averaged variational conditions given in Eq.~(\ref{eq:SA_variational}), the correspondence $\mathcal{L}_I
= E_I$ holds here. 
In the definition of the Lagrangian $\mathcal{L}_I$,
the parameters $\overline{\kappa}_{pq}^I$ and $\overline{\theta}_n^I$ are Lagrange multipliers designed to make it fully stationary such that
\begin{eqnarray}\label{eq:stationary-conditions}
\dfrac{\partial \mathcal{L}_I}{\partial \overline{\kappa}_{pq}^I}
=
\dfrac{\partial \mathcal{L}_I}{\partial \overline{\theta}_{n}^I}
=
\dfrac{\partial \mathcal{L}_I}{\partial \kappa_{pq}}
=
\dfrac{\partial \mathcal{L}_I}{\partial \theta_{n}} = 0.
\end{eqnarray}
To fulfil the stationary conditions in Eq.~(\ref{eq:stationary-conditions}), the Lagrange multipliers are determined by solving the so-called coupled-perturbed equations 
\begin{eqnarray}
\dfrac{\partial \mathcal{L}_I}{\partial \kappa_{rs}}
= \dfrac{\partial E_I}{\partial \kappa_{rs}}
+ \sum_{pq}
\overline{\kappa}_{pq}^I
H^{\rm OO}_{pq,rs}
+ \sum_n \overline{\theta}_n^I
H^{\rm CO}_{n,rs} = 0,\nonumber \\
\dfrac{\partial \mathcal{L}_I}{\partial \theta_m}
= \dfrac{\partial E_I}{\partial \theta_m} + \sum_{pq}\overline{\kappa}_{pq}^I
H^{\rm OC}_{pq,m}
+ \sum_n \overline{\theta}_n^I H^{\rm CC}_{n,m} = 0, \nonumber \\
\end{eqnarray}
where we have introduced 
\begin{align}
    H^{\rm OO}_{pq,rs}   &= \dfrac{\partial^2 E_\text{SA}}{\partial \kappa_{pq}\partial \kappa_{rs}},\\
    H^{\rm CC}_{n,m} &= \dfrac{\partial ^2 E_\text{SA}}{\partial \theta_{n} \partial \theta_{m}},\\
    H^{\rm CO}_{n,rs} &= \dfrac{\partial ^2 E_\text{SA}}{\partial \theta_{n} \partial \kappa_{rs}},
\end{align}
which correspond to matrix elements of the (state-averaged) orbital hessian $\bfH^{\rm OO}$, circuit hessian  $\bfH^{\rm CC}$ and circuit-orbital hessian $\bfH^{\rm CO}$ (with $\bfH^{\rm CO} = (\bfH^{\rm OC})^T$). The remaining  terms 
\begin{equation}
    G^{\text{O},I}_{rs} = \dfrac{\partial E_I}{\partial \kappa_{rs}} \quad \text{ and }  \quad
    G^{\text{C},I}_{m}  = \dfrac{\partial E_I}{\partial \theta_{m}},
\end{equation}
are elements of the circuit gradient vector $\mathbf{G}^{\text{C},I} $ and the orbital gradient vector $\mathbf{G}^{\text{O},I} $ of the state $| \Psi_I(\boldsymbol{\theta}) \rangle $. 
The orbital gradient for individual states $G^{{\rm O},I}_{pq}$ can be relatively easily computed from their one- and two-RDMs and MO coefficients.
The state-averaged orbital Hessian $H^{\rm OO}_{pq,rs}$
can also be determined from the state-averaged one- and two-RDMs and MO coefficients~\cite{staalring2001analytical,snyder2017direct}.
Therefore, they do not require any additional measurements on the quantum computer.
However, the first and second derivatives of the state-averaged energy with respect to the ans\"atze parameters requires many more measurements (but no additional qubits or deeper circuit depth). 
According to the parameter-shift rule~\cite{mitarai2018quantum}, $2^{2n}$ and $2^{4n}$ measurements are required to compute the gradient and Hessian with respect to ans\"atze parameters, respectively, for an ansatz with up to $n$-fold fermionic excitation operators. 
Note that $n=2$ is usually considered, as the trotterized-UCCSD ansatz can be made arbitrarily exact, as shown by Evangelista \textit{et al}~\cite{evangelista2019exact},
thus corresponding to 16 and 256 expectation values per element
of the $\mathbf{H}^{\rm CO}$ and $\mathbf{H}^{\rm CC}$ matrices, respectively.
Finally, the number of parameter-shifted RDMs to be measured on the quantum computer is directly related to the number of ans\"atze-parameters $N_p$.
For the circuit-orbital Hessian $\mathbf{H}^{\rm CO}$, $N_p$ parameter-shifted RDMs need to be measured, while $N_p (N_p + 1)/2$ are required to estimate the (symmetric) circuit-circuit Hessian matrix $\mathbf{H}^{\rm CC}$.
The total number of ans\"atze-parameters also scales
with the number of active orbitals as $\mathcal{O}(N_{\rm act}^4)$ for the generalized UCCD ansatz considered in this work.
For the sake of conciseness, 
we refer the interested reader to Appendix~\ref{app:GC_HCC} for more details about the estimation of the above Hessian matrices and gradients vectors.

Assuming we have evaluated the necessary Hessian matrices and gradient vectors out of a quantum circuit
following Appendix~\ref{app:GC_HCC}, the Lagrange multipliers $\overline{\kappa}_{pq}^I$ and $\overline{\theta}_n^I$ satisfying the conditions in Eq.~(\ref{eq:stationary-conditions}) are
determined on a classical computer by solving the following matrix equation
\begin{eqnarray}\label{eq:CPSAOOVQE_matrix}
\begin{pmatrix}
\bfH^{\rm OO} & \bfH^{\rm OC}\\
\bfH^{\rm CO} & \bfH^{\rm CC}
\end{pmatrix}
\begin{pmatrix}
\overline{\bmkappa}^I\\
\overline{\bmtheta}^I
\end{pmatrix}
=
- \begin{pmatrix}
\mathbf{G}^{\text{O},I} \\
\mathbf{G}^{\text{C},I}   
\end{pmatrix}.
\end{eqnarray}
Inserting these Lagrange multipliers back into Eq.~(\ref{eq:CP_SAOOVQE_energy_grad}) 
makes the Lagrangian fully stationary, and the property $\dv{E_I}{ x} = \dv{\mathcal{L}_I}{ x}= \pdv{\mathcal{L}_I}{ x}$ holds~\cite{helgaker1984second}. 
Hence, the energy derivative $\dv{E_I}{x} $ can be evaluated as follows:
\begin{equation} \label{eq:dEdx_core}
\begin{split}
\dfrac{d E_I}{d x} &= \sum_{pq} \pdv{h_{pq}}{x} \gamma_{pq}^{I,\rm eff}
+ \dfrac{1}{2}\sum_{pqrs} \pdv{ g_{pqrs}}{x} \Gamma_{pqrs}^{I,\rm eff}  \\
&  + \sum_J  \sum_n   w_J\bar{\theta}_n^I  G^{\text{C},J}_n(\tfrac{\partial \hat{\mathcal{H}}}{\partial x}),   
\end{split}
\end{equation}
with effective 1- and 2-RDMs defined by 
\begin{eqnarray}
{\bm \gamma}^{I,\rm eff} &=& {\bm \gamma}^I +  \tilde{\bm \gamma}^{I,\rm SA} \\
{\bm \Gamma}^{I,\rm eff} &=& {\bm \Gamma}^I + \tilde{\bm \Gamma}^{I,\rm SA}
\end{eqnarray}
with ${ \gamma}^I_{pq} = \bra{\Psi_I} \hat{E}_{pq} \ket{\Psi_I} $ and ${ \Gamma}^I_{pqrs} = \bra{\Psi_I} \hat{e}_{pqrs} \ket{\Psi_I} $ regular RDMs of the reference state $\ket{\Psi_I}$, supplemented by corrective state-averaged RDMs $\tilde{\bm \gamma}^{I,\rm SA}$ and $\tilde{\bm \Gamma}^{I,\rm SA}$ (encoding orbital contributions) with matrix elements  
\begin{eqnarray}\label{eq:orbital_contribution}
\tilde{\gamma}_{pq}^{I, \rm SA } =   \sum_o &\big( \gamma_{oq}^\text{SA} \overline{\kappa}_{op}^I + \gamma_{po}^{\rm SA} \overline{\kappa}_{oq}^I \big) \\
\tilde{\Gamma}_{pqrs}^{I, \rm SA } = \sum_o & \big( \Gamma^{\rm SA}_{oqrs} \overline{\kappa}_{op}^I + \Gamma_{pors}^{\rm SA}\overline{\kappa}_{oq}^I \nonumber \\
& + \Gamma^{\rm SA}_{pqos} \overline{\kappa}_{or}^I + \Gamma_{pqro}^{\rm SA} \overline{\kappa}_{os}^I \big) .
\end{eqnarray}
Note that building these effective matrices does not require any additional measures from the quantum circuit as the RDMs ${\bm \gamma}^I$ and ${\bm \Gamma}^I$ are already evaluated during the SA-OO-VQE to estimate the state-averaged energy.  The circuit gradient $G^{\text{C},J}(\frac{\partial \hat{\mathcal{H}}}{\partial x})$ introduced in Eq.~(\ref{eq:dEdx_core}) is defined such that
\begin{eqnarray}
G^{\text{C},J}_n(\tfrac{\partial \hat{\mathcal{H}}}{\partial x}) =  \frac{\partial }{\partial \theta_{n}}   \bra{\Psi_J} \frac{\partial  \hat{\mathcal{H}} }{\partial x}   \ket{\Psi_J}  
\end{eqnarray}
and can be estimated out of a quantum circuit in the same way as for a generic energy gradient using for example the parameter-shift rule (cf Appendix \ref{app:GC_HCC}). The change being here that the central operator is now the nuclear derivative of the Hamiltonian $ \partial  \hat{\mathcal{H}} /  \partial x$  which can be evaluated on a classical computer as shown in Appendix~\ref{app:derivative_hamiltonian}.
Note that, we also refer the interested reader to Appendix~\ref{app:derivative_hamiltonian} for practical details about nuclear derivatives of electronic integrals (in Eq.~(\ref{eq:dEdx_core})) which can be evaluated on a classical computer with common quantum chemistry packages.

Interestingly, compared to its classical analogue SA-CASSCF,
note that a unique set of ans\"atze parameters $\bmtheta$ is considered to simultaneously find both ground and first excited states in SA-OO-VQE, instead of the configuration-interaction (CI) coefficients for each state (denoted by $\mathbf{c_0}$ and $\mathbf{c_1}$).
This results in a much reduced size of the parameter space,
\begin{eqnarray}
\dim(\boldsymbol \theta) \ll \dim(\mathbf{c}_0) + \dim(\mathbf{c}_1).
\end{eqnarray}
This has important consequences, as the original CP-MCSCF equations sometimes cannot be solved due to memory issues in storing all the matrix elements of $\mathbf{H}^{\rm CC}$, although some alternative implementations have been proposed to overcome this problem (see Ref.~\citenum{snyder2017direct} and references therein).
Hence, the classical complexity in solving the coupled-perturbed equations [Eq.~(\ref{eq:CPSAOOVQE_matrix})] is considerably reduced in SA-OO-VQE compared to SA-CASSCF,
at the expense of a lower accuracy (as the SA-VQE solver is not exact in contrast to SA-CASSCF).

\subsubsection{Non-adiabatic couplings}\label{subsubsec:NAC}
  
Non-adiabatic couplings have been calculated recently by Tamiya {\it et al.}~\cite{tamiya2021calculating}
in the context of SS-VQE without any orbital optimization.
In this work, we provide an analytical approach to estimate non-adiabatic couplings within the SA-OO-VQE algorithm, for which the state-averaged orbital-optimization procedure implies a more involved derivation.
The definition and Hellmann--Feynman formula for the NAC, $D_{IJ}$, have been given above -- see Eq.~(\ref{eq:def_NAC}) and Eq.~(\ref{eq:form_NAC_HF}) -- in the ideal case of exact adiabatic eigenstates. 
It is well-known in the practical context of an MCSCF ansatz that this term actually splits into two contributions: (i) a typically larger CI-contribution, which obeys a Hellmann--Feynman like formula (except that eigenstates are now CI-coefficient vectors and the Hamiltonian operator is replaced by its finite matrix representation in the CSF basis set); 
(ii) a typically smaller CSF-contribution, which accounts for molecular orbital gradients (via both their expansion coefficients and the overlaps among the primitive atomic basis functions); see, e.g., Ref.~\citenum{lengsfield1992nonadiabatic}. 
While the latter CSF contribution is usually straightforward to estimate, the former CI contribution is a more involved term which should take into account the non-variational character of MCSCF wavefunctions. 
Fortunately, coupled-perturbed equations have been derived to treat this aspect based on the same machinery as for gradient calculation~\cite{lengsfield1984evaluation, staalring2001analytical,lengsfield1992nonadiabatic,yarkony1995modern,snyder2015atomic,snyder2017direct, fdez2016analytical}. 
We employed a similar approach to obtain an analytical estimation of NACs with SA-OO-VQE wavefunctions. 
For sake of conciseness, we will present in the following only the essential equations of our developments (we refer the interested reader to
Appendix~\ref{app:NAC_deriv} where we detail each step of the derivation). In practice, one has to solve the following set of coupled linear equations 
\begin{eqnarray}\label{eq:Lagrange_mult_matrix_NAC}
\begin{pmatrix}
\bfH^{\rm OO} & \bfH^{\rm OC}\\
\bfH^{\rm CO} & \bfH^{\rm CC}
\end{pmatrix}
\begin{pmatrix}
\overline{\bmkappa}^{IJ}\\
\overline{\bmtheta}^{IJ}
\end{pmatrix}
=
- \begin{pmatrix}
\mathbf{G}^{\text{O},IJ}\\
 0
\end{pmatrix},
\end{eqnarray}
to determine the NAC Lagrange multipliers $\overline{\bmkappa}^{IJ}$ and
$\overline{\bmtheta}^{IJ}$. 
In Eq.~(\ref{eq:Lagrange_mult_matrix_NAC}), we retrieve the same Hessian blocks as for the gradient calculation and $G^{\text{O},IJ}_{pq} = \bra{\Psi_I} (\partial \mathcal{H}/ \partial\kappa_{pq}) \ket{\Psi_J}$ represents the interstate orbital coupling gradient whose elements can be easily measured out of a quantum computer (using for example methods provided in Ref.~\citenum{nakanishi2019subspace}). 
Once the multipliers are determined using a classical computer, the NAC can be evaluated as follows:
\begin{equation}\label{eq:NAC_final}
\begin{split}
    D_{IJ} = \frac{1}{ E_J - E_I } \Bigg( & \sum_{pq} \pdv{h_{pq}}{x} \gamma_{pq}^{IJ,\rm eff} + \dfrac{1}{2}\sum_{pqrs} \pdv{ g_{pqrs}}{x} \Gamma_{pqrs}^{IJ,\rm eff}  \\ 
 + &\sum_K  \sum_n   w_K \overline{\theta}_n^{IJ}   G^{\text{C},K}_n(\tfrac{\partial \hat{\mathcal{H}}}{\partial x}) \Bigg)\\
 - &\dfrac{1}{2}\sum_{pq}\gamma_{pq}^{IJ} \big( (\partial_x p|q) - (q | \partial_x p) \big). 
\end{split}
\end{equation}
The effective transition 1- and 2-RDMs introduced here are defined by
\begin{eqnarray}
{\bm \gamma}^{IJ,\rm eff} &=& {\bm \gamma}^{IJ}+  \tilde{\bm \gamma}^{IJ,\rm SA} \\
{\bm \Gamma}^{IJ,\rm eff} &=& {\bm \Gamma}^{IJ} + \tilde{\bm \Gamma}^{IJ,\rm SA} ,
\end{eqnarray}
and $\tilde{\bm \gamma}^{IJ,\rm SA}$ and $\tilde{\bm \Gamma}^{IJ,\rm SA}$ are the orbital contributions to the 1- and 2-RDMs, respectively. These state-averaged matrices are defined in a same way as in Eq.~(\ref{eq:orbital_contribution}) (where we replace $\overline{\kappa}^{I}$ by $\overline{\kappa}^{IJ}$). 
In Eq.~(\ref{eq:NAC_final}), the terms in parentheses encode the off-diagonal Hellmann--Feynman contribution complemented by additional corrective terms accounting for the non-variational character of the wavefunctions due to orbital and quantum circuit optimization. 
The contribution outside parenthesis is the so-called ``CSF-term'' which formally takes into account the variation of the Slater determinants due to nuclear displacement (see Ref.~\citenum{lengsfield1984evaluation} for more details).  
The elements $(\partial_x p|q)$ represent the half-derivative of MOs' overlap which can be easily calculated analytically with most quantum chemistry packages.
Note the presence of transition 1- and 2-RDMs in Eq.~(\ref{eq:NAC_final}) defined as $\gamma_{pq}^{IJ} = \bra{\Psi_I} \hat{E}_{pq} \ket{\Psi_J}$ and $\Gamma_{pqrs}^{IJ} = \bra{\Psi_I} \hat{e}_{pqrs} \ket{\Psi_J}$. These matrices can be obtained from a quantum circuit using methods to determine transition matrix elements such as the one provided in Ref.~\citenum{nakanishi2019subspace}. 

\section{Computational details}

To test our theoretical developments, we consider the formaldimine molecule $\text{CH$_2$NH}$, a minimal Schiff base model relevant for the study of the  photoisomerization in larger bio-molecules (such as the RPSB molecule whose \textit{cis} to \textit{trans} isomerization plays a key role in the visual cycle process~\cite{chahre1985trigger,birge1990nature,schnedermann2018evidence}). 
An illustration of the geometry of the molecule is shown in Fig.~\ref{fig:illustration_SAOOVQE}a. 

In analogy with our previous study~\cite{Yalouz_2021}, we freeze and constrain the N--CH$_2$ part of the molecule in the same plane.
The interatomic distances are $d_\text{N--C}=1.498~\text{\AA}$, $d_\text{C--H}=1.067~\text{\AA}$, $d_\text{N--H}=0.987~\text{\AA}$  and the angle $\widehat{\text{N--C--H}} = 118.36^\circ$. 
The second H atom is symmetric to the first one with respect to the N--C axis.
The two remaining degrees of freedom characterize the out-of-plane bending angle $\alpha \equiv \widehat{\text{H--N--C}}$ and the dihedral angle $\phi \equiv \widehat{\text{H--N--C--H}}$.
For practical calculations, the cc-pVDZ basis is used and an active space of four electrons in three orbitals (4,3) is considered.
The orbital optimization is realized over the 43 spatial-orbitals of the system (for SA-OO-VQE and SA-CASSCF).
Reference quantum chemistry calculations are realized with OpenMolcas~\cite{OPEN_MOLCAS} (e.g. SA-CASSCF simulation and estimation of the associated gradients and NAC) whereas the Psi4~\cite{smith2020psi4} package is used to provide SA-OO-VQE with initial data about the molecular system.

The noiseless state-vector simulation of the SA-OO-VQE algorithm is realized using the python quantum computing packages OpenFermion~\cite{mcclean2020openfermion} and Cirq~\cite{cirq}. The ansatz we employ in the SA-VQE algorithm is a
generalized unitary coupled cluster ansatz with spin-free double-excitation operators (GUCCD) such that
\begin{align}
&\hat{U}(\boldsymbol{\theta})=e^{\hat{T}(\boldsymbol{\theta})-\hat{T}^{\dag}(\boldsymbol{\theta})},\\
&\hat{T}(\boldsymbol{\theta}) = \sum_{t, v, w, u}^\text{active} \theta_{tuvw} \sum_{\sigma,\tau=\uparrow,\downarrow} \hat{a}^\dagger_{t\sigma} \hat{a}^\dagger_{v\tau} \hat{a}_{w\tau} \hat{a}_{u\sigma}.
\label{eq:GCC}
\end{align}
Our simulation considers ans\"atze parameters $\boldsymbol{\theta}$ initialized to zero and optimized using the ``Sequential Least Squares Programming'' (SLSQP) method from the python Scipy package.
For each call of SA-VQE, the SLSQP method is run with a maximum number of 500 iterations and a precision threshold of $10^{-8}$ Ha.
The threshold for the global convergence of SA-OO-VQE is also set to $10^{-8}$~Ha. A home-made python code has been developed
to implement the state-averaged Newton-Raphson algorithm required for the SA-OO subalgorithm.
For an active space of four electrons in three orbitals, we have a circuit depth of 2688 (+6 for the initial rotation circuit), and we optimize 12 parameters.
We refer the interested reader to our previous work~\cite{Yalouz_2021} for more details about the GUCCD ansatz (such as gate complexity).

\section{Numerical results}\label{sec:results}

\subsection{Illustration of the final state resolution in SA-OO-VQE}\label{subsec:illustration_stateresolution}

\begin{figure}
\includegraphics[width=0.7\columnwidth]{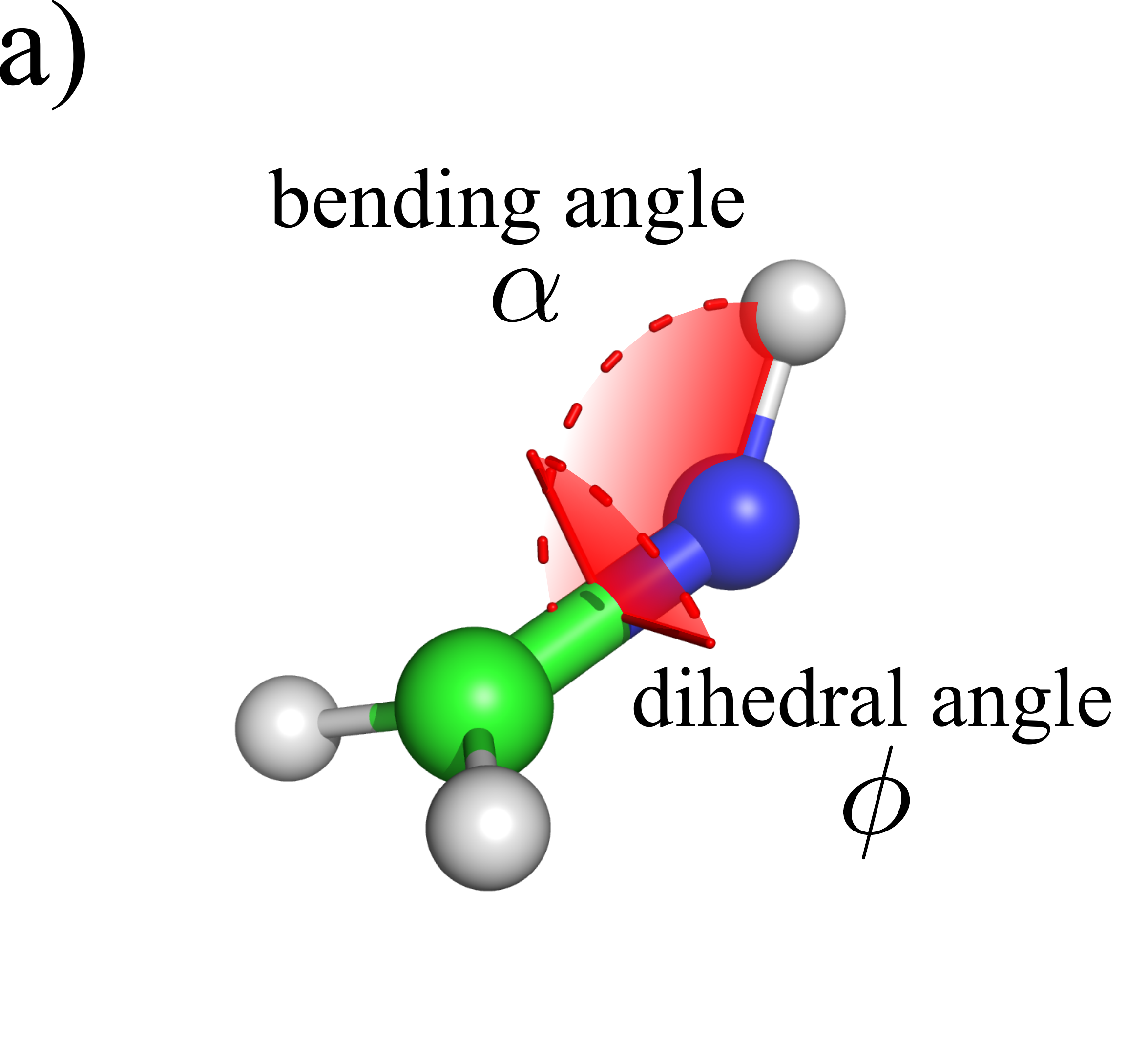} 
\includegraphics[width=\columnwidth]{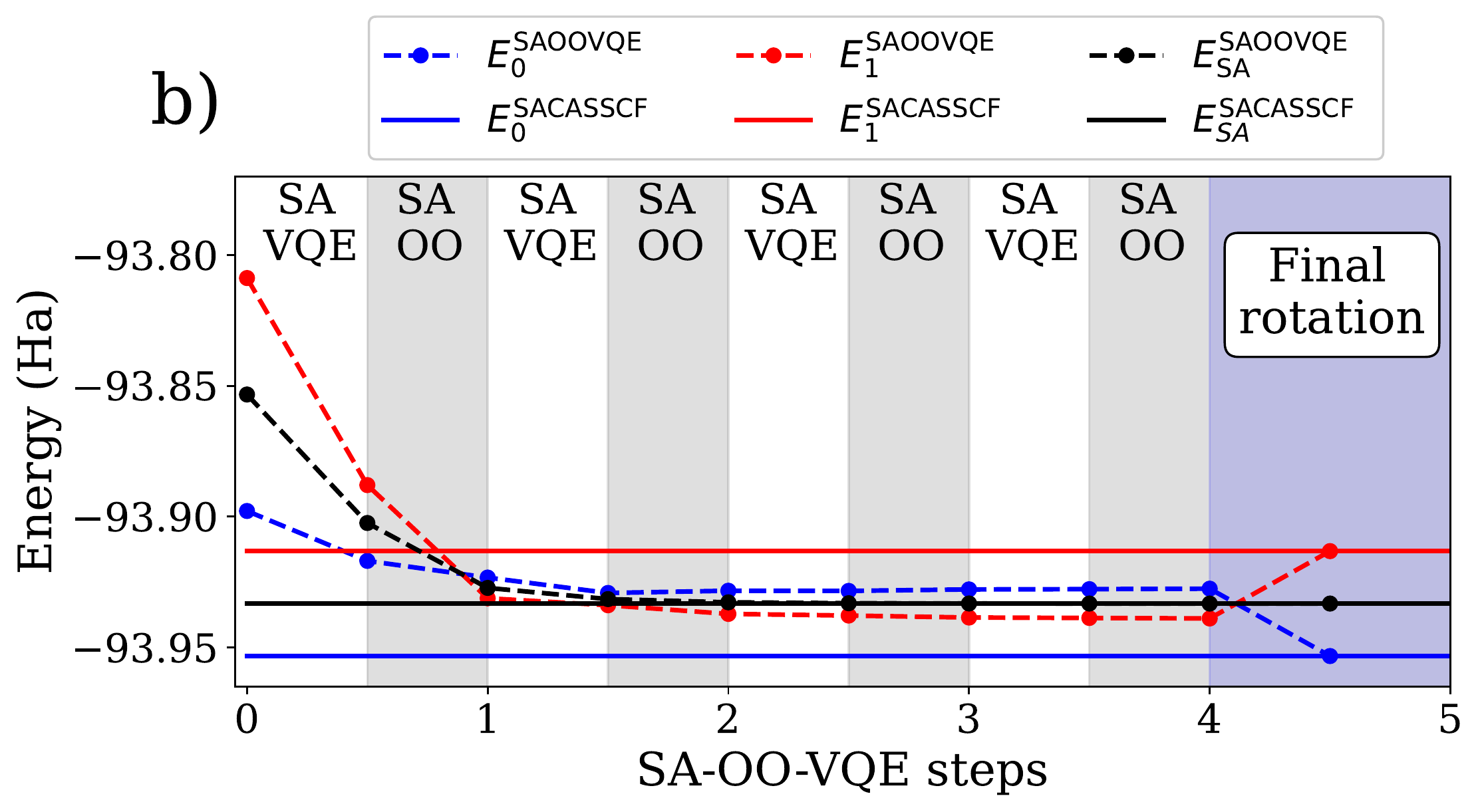}  
\includegraphics[width=\columnwidth]{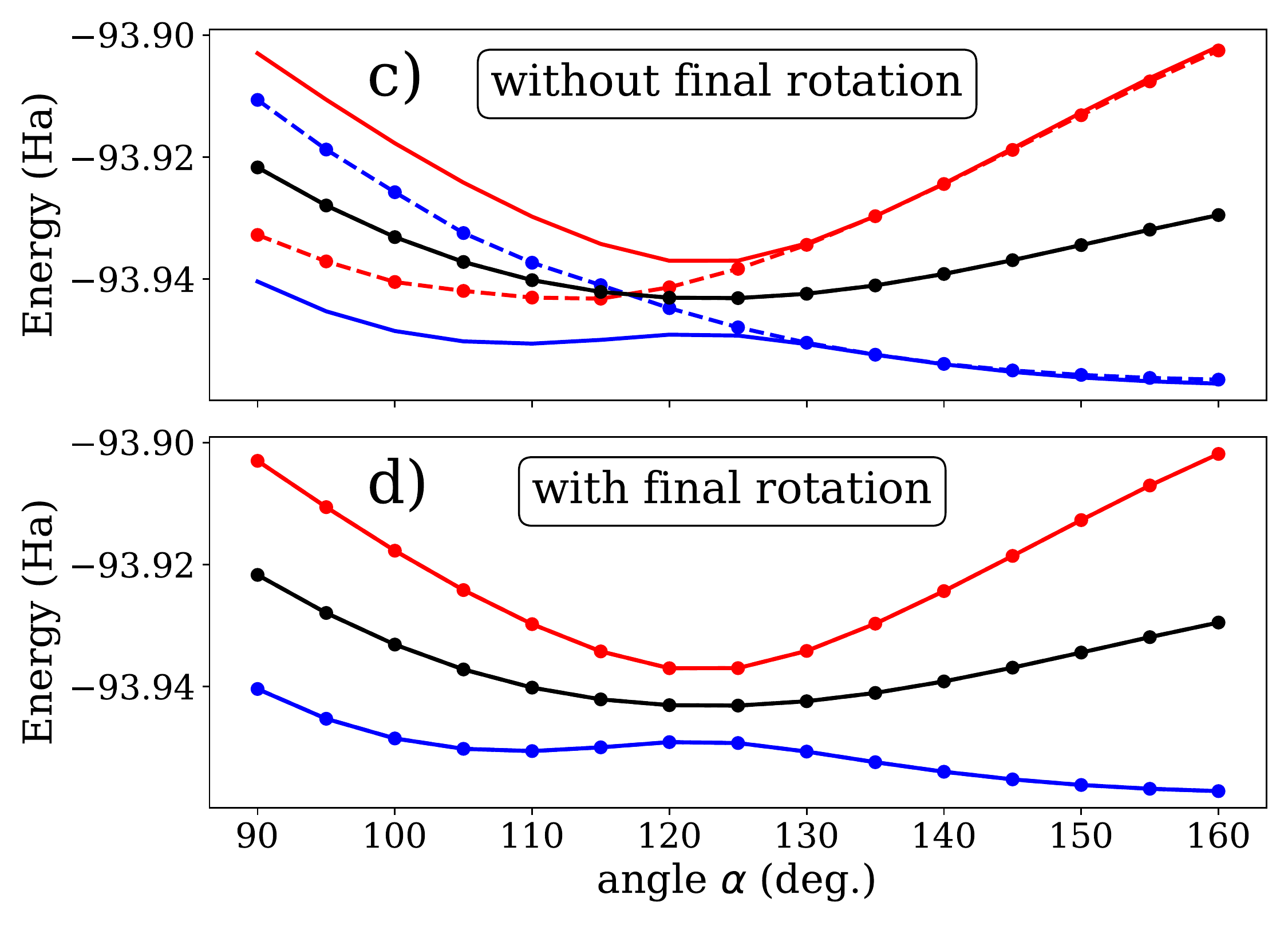}
\caption{\textbf{Illustration of the convergence
of each step of the equi-ensemble SA-OO-VQE algorithm}.
In all panels, dashed and solid lines represent the SA-OO-VQE and SA-CASSCF energies, respectively.
\textbf{a)} Geometry of the formaldimine molecule with the bending angle $\alpha = \widehat{\text{NCH}}$ and the dihedral angle $\phi=\widehat{\text{HCNH}}$.
\textbf{b)} Evolution of the state-averaged energy during the different steps of the SA-OO-VQE algorithm
for $\phi=85^\circ$ and $\alpha=100^\circ$. 
White and grey strips represent SA-VQE and SA-OO phases, respectively, while the blue strip represents the final state-resolution step. 
Converged SA-OO-VQE 1D-PES scans along $\alpha$ with $\phi=85^\circ$ are shown before (\textbf{c}) and after (\textbf{d}) the state-resolution procedure.}
\label{fig:illustration_SAOOVQE}
\end{figure}

\begin{figure*}
\centering
\resizebox{\textwidth}{!}{
    \includegraphics{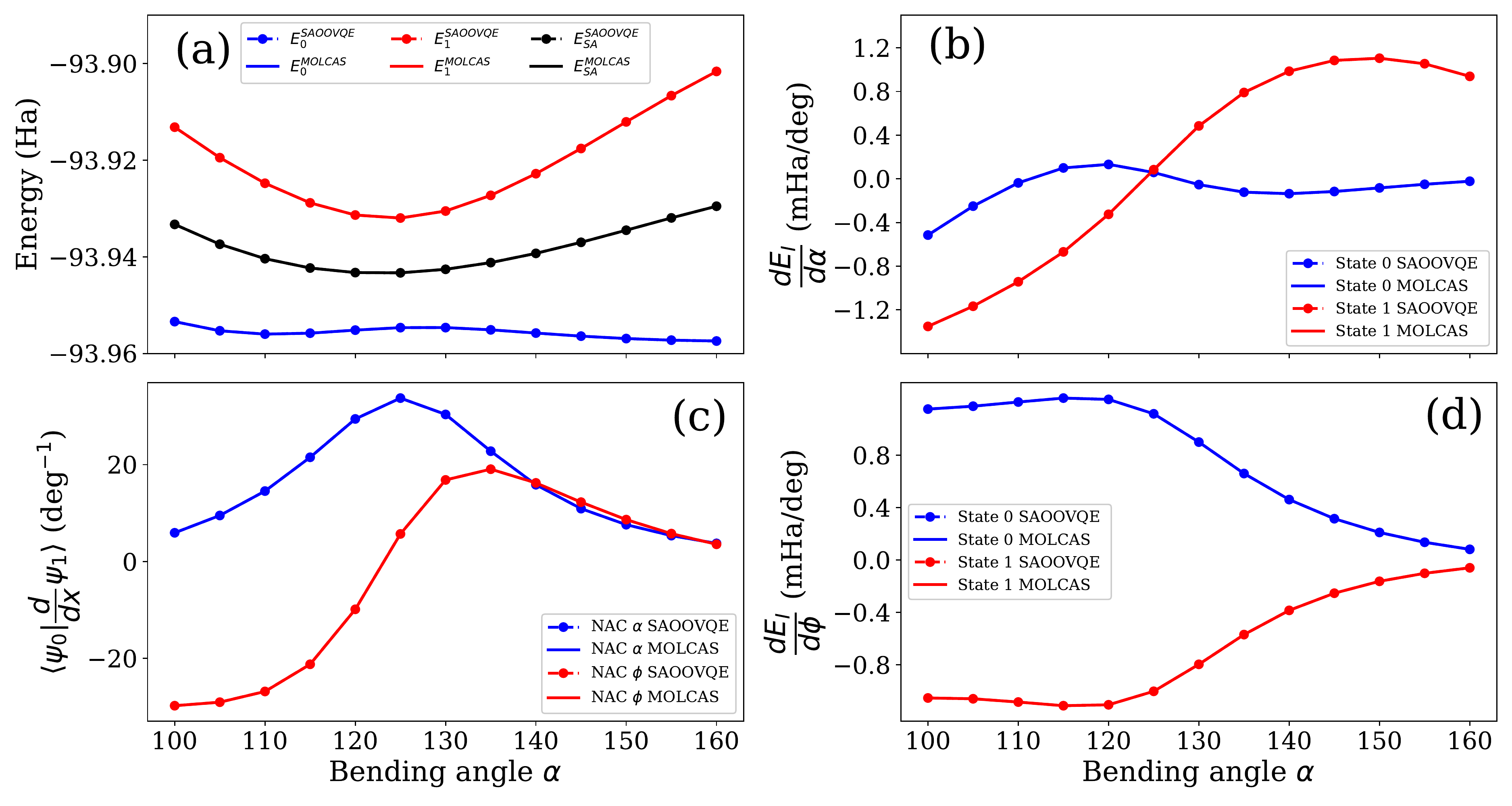}
    }
    \caption{\textbf{Energy gradients and NAC with respect to the bending angle $\alpha$ with $\phi=80^\circ$}. 
    SA-OO-VQE results are shown in dashed lines with dots while solid lines are for the reference SA-CASSCF results.
    The ground and first-excited state energies
    are represented by blue and red colors, respectively.
    \textbf{(a)} Potential energy surfaces
    \textbf{(b)} Analytical individual-state energy gradients [Eq.~(\ref{eq:dEdx_core}) for $x = \alpha$]
    \textbf{(c)} non-adiabatic coupling vector $\psh{\Psi_0}{(\partial/\partial x)\Psi_1}$ for $x = \alpha$ (blue) and $x= \phi$ (red)
    \textbf{(d)} Analytical individual-state energy gradients [Eq.~(\ref{eq:dEdx_core}) for $x = \phi$].}
    \label{fig:formaldimine_all_80}
\end{figure*}

As discussed in Sec.~\ref{subsubsec:control},
the SA-OO-VQE states
obtained after optimizing the $\bmkappa$ and $\bmtheta$
parameters do not correspond to the eigenstates of $\hat{\mathcal{H}}(\bmkappa)$, and an additional rotation between the initial states is required (also called state resolution).
In Fig.~\ref{fig:illustration_SAOOVQE},
we illustrate the convergence of each step of the SA-OO-VQE algorithm applied to the formaldimine molecule,
depicted in panel (a), with geometry parameters set to $\phi=80^\circ$ and $\alpha=100^\circ$.
The ground, first-excited and state-averaged energies
are plotted on panel (b) for each step of the SA-OO-VQE algorithm, where SA-VQE and SA-OO phases
are represented by white and grey strips, respectively.
The ground and first-excited SA-CASSCF reference energies
are also provided for comparison, as well as
the state-averaged SA-CASSCF energy which forms a natural lower bound for SA-OO-VQE [see Eq.~(\ref{eq:SAOOVQE_energy})].
As readily seen in panel~(b), alternating between the SA-VQE and the SA-OO algorithms progressively lowers the state-averaged energy, requiring three full SA-OO-VQE cycles to reach global convergence.
At convergence, this energy has an error of only
$\sim 10^{-6}$ Ha with respect to SA-CASSCF,
indicating that the subspace spanned by the SA-OO-VQE trial states is a very good approximation to the one spanned by the SA-CASSCF states.
However, the converged individual SA-OO-VQE states
differ significantly from the SA-CASSCF states (at the end of the third SA-OO-VQE step in panel (b)).
Hence, one has to apply to state resolution such as
described in Sec.~\ref{subsubsec:control}
to recover the correct eigenstates.
This final step is symbolized by the blue region in panel (b), where we employ the rotation circuit shown in Fig.~\ref{fig:resolve} and optimize the rotation parameter $\varphi$
such that the energies are effectively pushed as far as possible from each other, thus maximizing the difference between the first-excited and ground-state energies.
After this final step,
the individual SA-OO-VQE energies are in very good agreement with the SA-CASSCF ones (with an error of $ \sim 10^{-6}$ Ha, similar to the state-averaged energy error).
Note that such agreement is expected, as single and double excitations are enough to span all the electronic configurations in the case of an active space (4,3).
Deviations from the SA-CASSCF results may appear when considering larger active spaces.

In panels (c) and (d) of Fig.~\ref{fig:illustration_SAOOVQE}, we show the one-dimensional (1D) PES along the $\alpha$-angle
for
SA-OO-VQE and SA-CASSCF with a dihedral angle $\phi=85^\circ$.
More precisely, we compare the 1D-PES of SA-OO-VQE without (panel (c)) and with (panel (d)) the final state-resolution procedure. 
As readily seen in these two panels, the state-averaged SA-OO-VQE energy is in very good agreement with the reference state-averaged SA-CASSCF energy  all over the 1D-PES (with an error $ < 10^{-6}$ Ha). 
Without state resolution (see panel (c)), the individual-state energies are globally different from the SA-CASSCF ones, especially for $\alpha < 130^\circ$.
For $\alpha > 130^\circ$, the energies of the individual states match the reference ones, showing that the state resolution is not always necessary to capture the eigenstates.
Interestingly, the SA-OO-VQE states smoothly cross around $\alpha \sim 118^\circ$.
As discussed in Ref.~\citenum{Yalouz_2021},
this results from the use of an equi-ensemble
where no ordering of the trial states is enforced.
In such a case, the converged SA-OO-VQE states
will naturally evolve to the state that is the closest to its initial state, \textit{i.e.} with the highest overlap with its initial state (see Ref.~\citenum{Yalouz_2021} for more details).
Typically, when $\alpha < 118^\circ$ the singlet single-excited CIS state $\ket{\Phi_B} = -\hat{E}_\text{hl}\ket{\text{HF}}$ has the largest contribution to the SA-CASSCF ground state, while
for $\alpha > 132^\circ$ the $\ket{\Phi_A} = \ket{\text{HF}}$ has the largest contribution (and reciprocally for the first-excited SA-CASSCF state).
The state-resolution procedure (see panel (d)) will lift the crossing, thus resulting in an avoided-crossing captured by the adiabatic eigenstates and an excellent agreement between the SA-OO-VQE and SA-CASSCF energies.

\begin{figure*}
\resizebox{\textwidth}{!}{
    \includegraphics{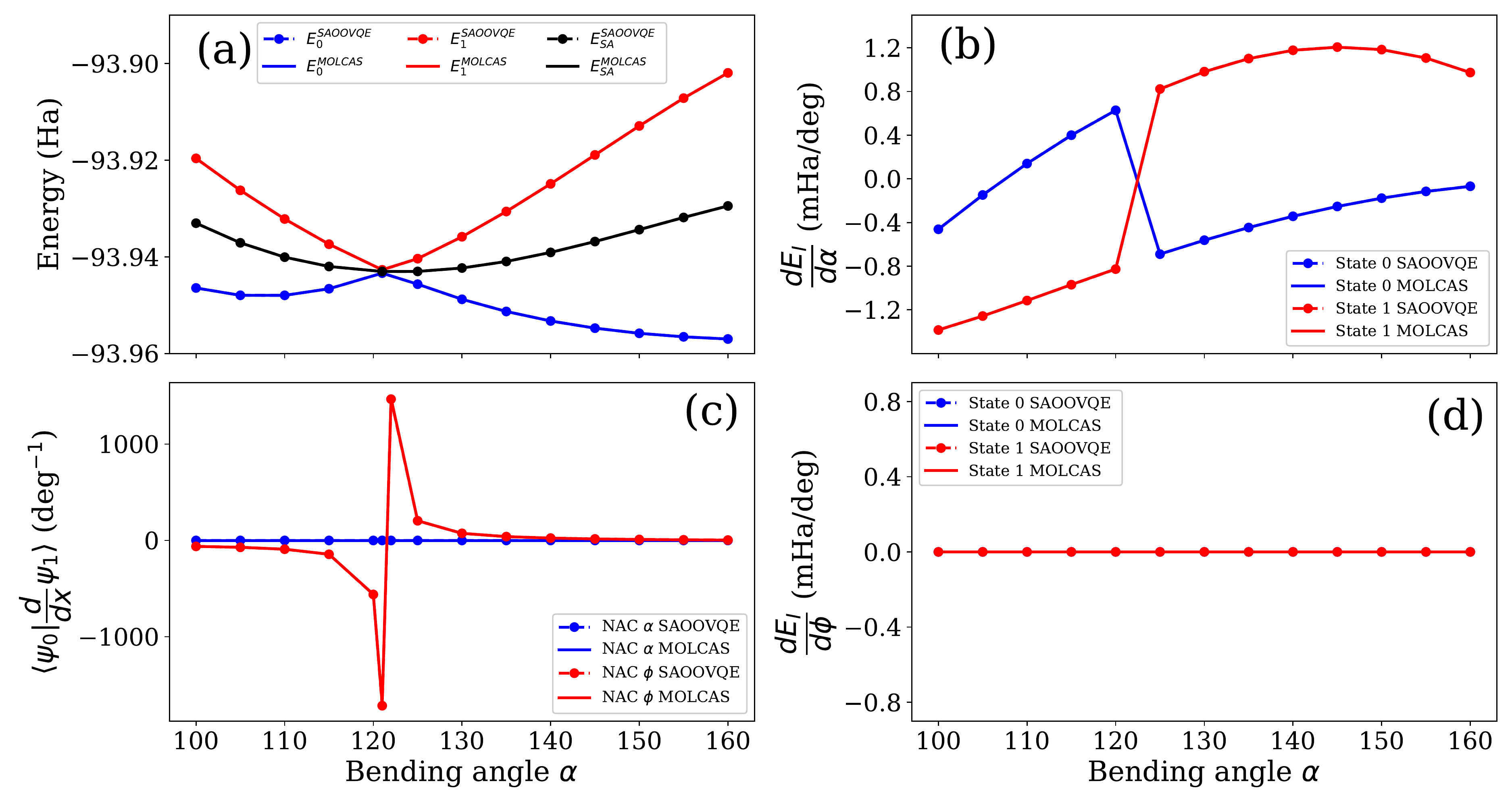}
    }
    \caption{\textbf{Energy gradients and NAC with respect to the bending angle $\alpha$ with $\phi=90^\circ$}.
    SA-OO-VQE results are shown in dashed lines with dots while solid lines are for the reference SA-CASSCF results.
    The ground and first-excited state energies
    are represented by blue and red colors, respectively.
    \textbf{(a)} Potential energy surfaces
    \textbf{(b)} Analytical individual-state energy gradients [Eq.~(\ref{eq:dEdx_core}) for $x = \alpha$]
    \textbf{(c)} non-adiabatic coupling vector $\psh{\Psi_0}{(\partial/\partial x)\Psi_1}$ for $x = \alpha$ (blue) and $x= \phi$ (red)
    \textbf{(d)} Analytical individual-state energy gradients [Eq.~(\ref{eq:dEdx_core}) for $x = \phi$].
    \label{fig:formaldimine_all_90}}
\end{figure*}

\subsection{Calculation of analytical gradients and non-adiabatic couplings}\label{subsec:gradients_NAC_results}

Let us now turn to the nuclear gradients and non-adiabatic couplings of the adiabatic states obtained after convergence of the SA-OO-VQE algorithm (with state resolution), following Eqs.~(\ref{eq:dEdx_core})
and (\ref{eq:NAC_final}). As mentioned in section~\ref{subsubsec:analytical_gradients}, the extra cost for the quantum device is in determining the Hessians $\mathbf{H}^{\rm CC}$ and $\mathbf{H}^{\rm OC}$. As we have 12 parameters in our simulation, this amounts to 78  and 12 entries of $\mathbf{H}^{\rm CC}$ and $\mathbf{H}^{\rm OC}$ to be computed, respectively. With the parameter shift rule, this becomes a total amount of 19968 and 3072 measurements for $\mathbf{H}^{\rm CC}$ and $\mathbf{H}^{\rm OC}$, respectively. The rest of the computational work is done on a classical device, given the Hessians and the (transition) 1- and 2-RDM to compute the analytical gradient (non-adiabatic coupling). 
The results are represented in Figs.~\ref{fig:formaldimine_all_80} and ~\ref{fig:formaldimine_all_90}
 along the $\alpha$ direction for $\phi = 80^\circ$ and $\phi = 90^\circ$, respectively.
As readily seen in Figs.~\ref{fig:formaldimine_all_80} and ~\ref{fig:formaldimine_all_90},
both the analytical gradients and the NAC
calculated from our
SA-OO-VQE implementation cannot be distinguished from the SA-CASSCF results, with
a negligible difference of the order of $10^{-3}$ mHa/degree for the gradients and $10^{-3}$ degree$^{-1}$ for the NAC amplitudes (all over the PES).
This supports the derivations of Eqs.~(\ref{eq:dEdx_core})
and (\ref{eq:NAC_final}) and shows that SA-OO-VQE can provide (ideally, \textit{i.e.} without noise) as accurate results as its classical SA-CASSCF analogue.
Turning to the energy landscape of Fig.~\ref{fig:formaldimine_all_80} (panel~(a)) with $\phi=80^\circ$
(exactly the same as Fig.~\ref{fig:illustration_SAOOVQE}, panel~(d), but plotted again here for convenience),
we observe an avoided crossing between the ground and first-excited states around $\alpha = 125^\circ$.
This particular behavior can also be detected by looking at the amplitudes of the NAC (panel~(c) of Fig.~\ref{fig:formaldimine_all_80}), which increase significantly at the avoided crossing position (but without diverging).

\begin{figure}[!ht]  
\includegraphics[width=\columnwidth]{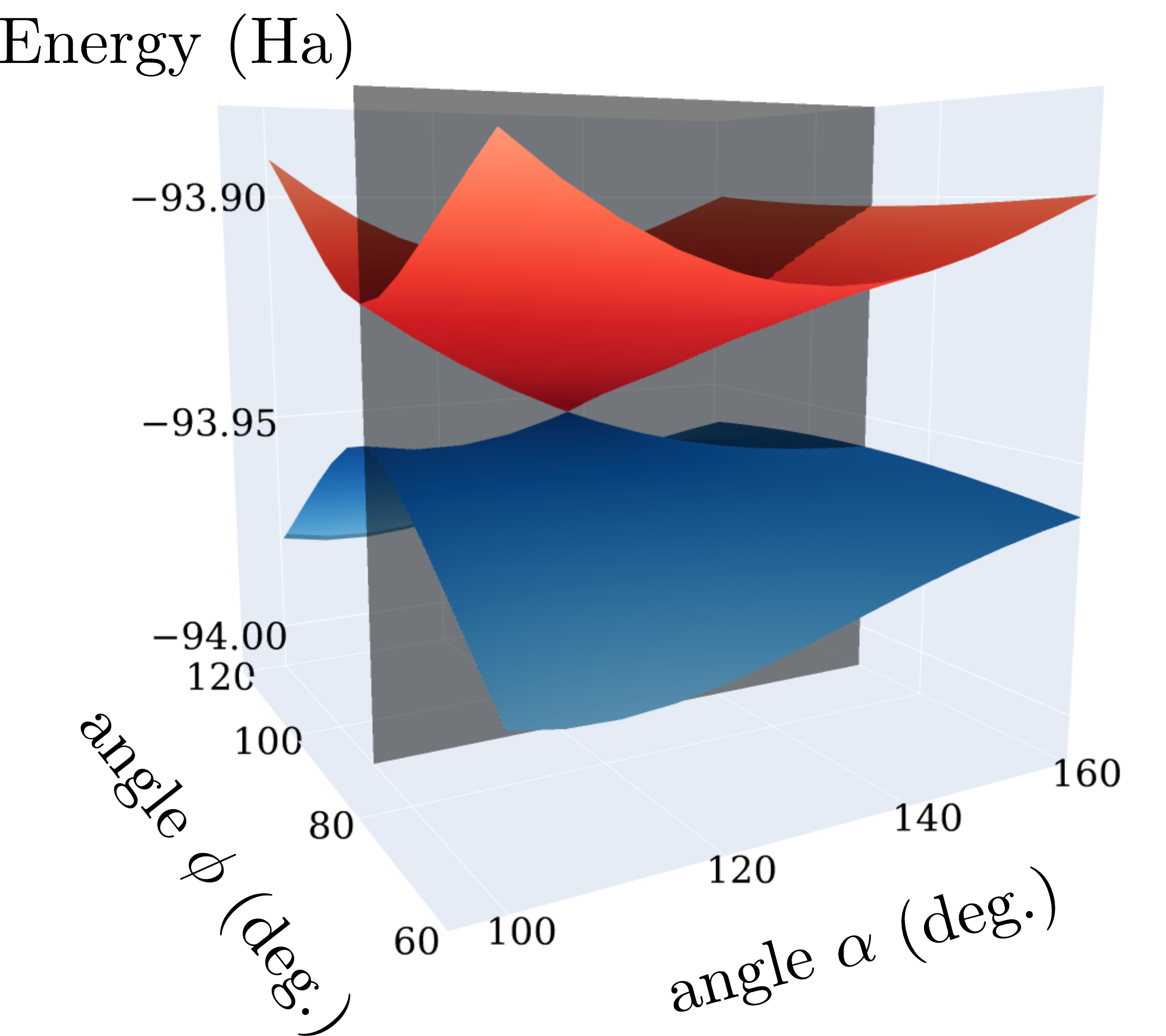} 
\caption{ \textbf{$(\alpha,\phi)$-PES for the formaldimine molecule.} The energies are obtained with the SA-OO-VQE algorithm after a full resolution of the states. A conical intersection is observed around the geometry $(\alpha = 121.5^\circ, \phi=90^\circ)$. Grey plane is defined for $\phi=90^\circ$ and intersects both PESs at extrema values in the $\phi$ direction.}
\label{fig:3D}
\end{figure}

\begin{figure}
\resizebox{\columnwidth}{!}{
    \includegraphics{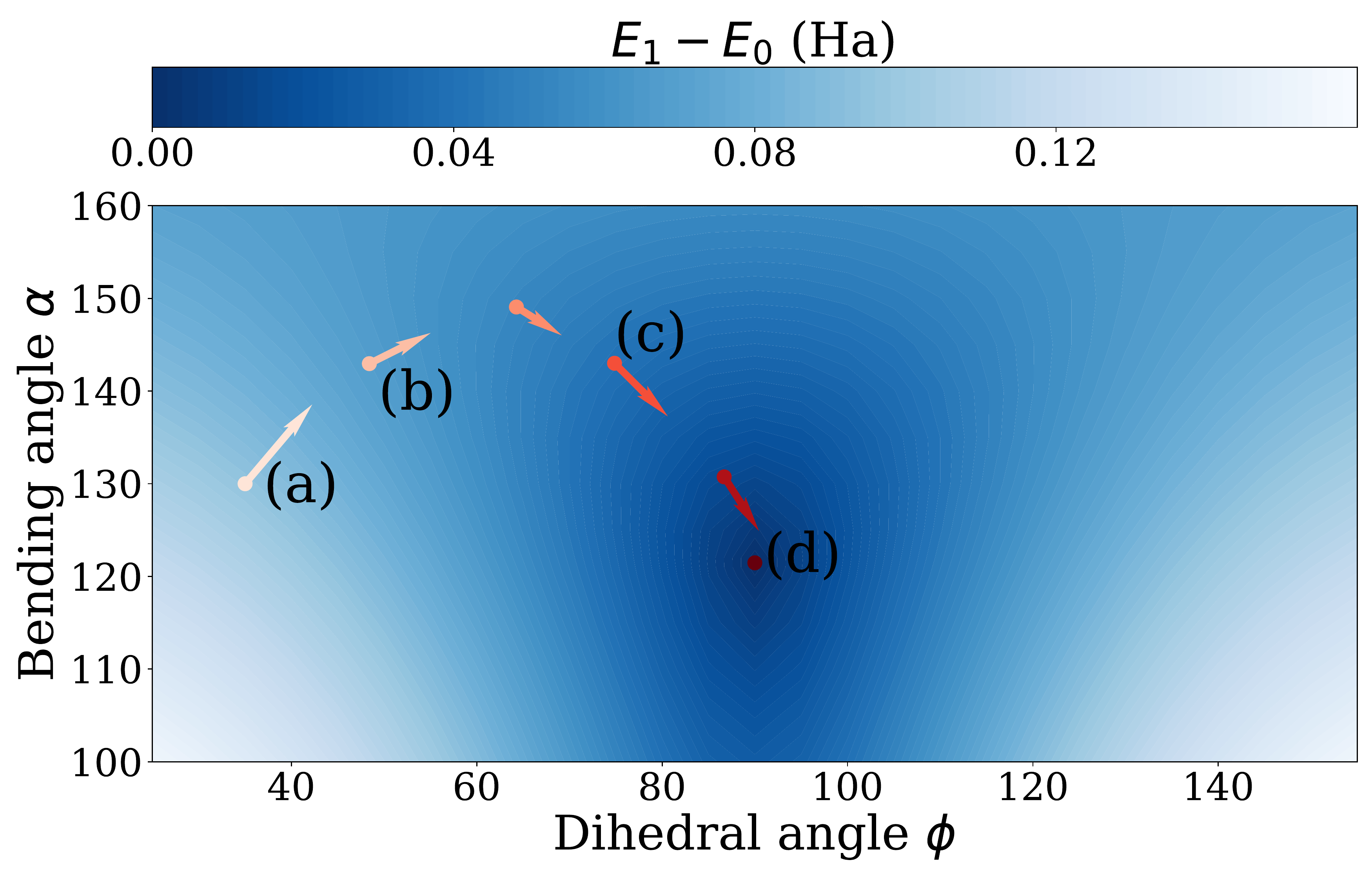}
    }
\resizebox{\columnwidth}{!}{
    \includegraphics{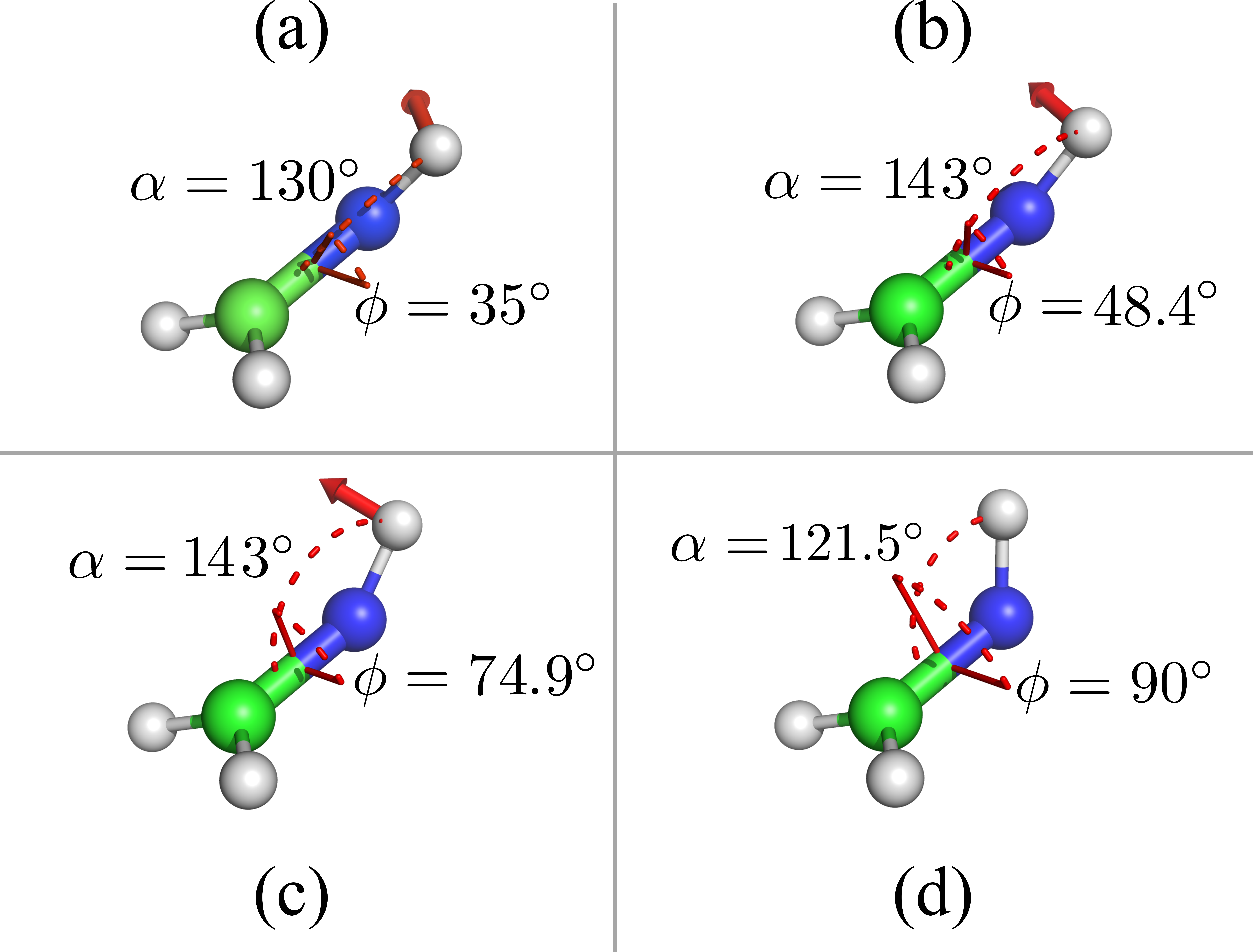}
    }
\caption{\textbf{$\boldsymbol{(\alpha,\phi)}$-Geometry optimization to the conical intersection point}.
\textbf{Left panel:} optimization path of a steepest-descent algorithm to locate the conical intersection of formaldimine. 
The contourplot shows the energy difference $\Delta E = E_1-E_0$, and the vectors represent the negative of the gradient of the energy difference $\mathbf{g}_{\Delta E}$ at each point of the optimization.
\textbf{Right panel:} Four molecular geometries corresponding to four points of the optimization, denoted by (a), (b), (c) and (d) on the left panel. The corresponding $(\alpha,\phi)$ angles and $\mathbf{g}_{\Delta E}$ vectors are shown.
\label{fig:CI_optimization}}
\end{figure}

In contrast, for $\phi=90^\circ$
(see panel~(a) of Fig.~\ref{fig:formaldimine_all_90}) the 1D-PES shows a crossing (a conical intersection here).
This very different behaviour is manifested
by a discontinuity in the gradients $\partial E_0/ \partial \alpha$ and $\partial E_1/ \partial \alpha$  that suddenly inverse their position at the crossing point $\alpha \approx 121.5^\circ$,
while the gradients are smoothly evolving along the 1D-PES when the states do not cross (see panel~(b) of Figs.~\ref{fig:formaldimine_all_80} and \ref{fig:formaldimine_all_90}, respectively).
This is the direct consequence of the presence of a degeneracy in the energy profile.
In contrast to the $\alpha$ direction (panel~(b)), the fact that the gradients $ \partial E_0 / \partial \phi$ and $ \partial E_1 / \partial \phi$ are zero for all $\alpha$ 
(see panel~(d) in Fig.~\ref{fig:formaldimine_all_90})
reveals the presence of extrema for both states in the $\phi$ direction. 
This behaviour is consistent with $\phi = 90^\circ$ defining a mirror-plane symmetry $\sigma_v$
($\mathcal{C}_{\rm s}$ point group)
wherein there is no interstate coupling (different irreducible representations) and local extrema are induced for both potential energies at $\phi = 90^\circ$.
For a better illustration, these local extrema along the $\phi$ direction are shown in Fig.~\ref{fig:3D} and correspond to the intersection points between the grey plane (defining $\phi=90^\circ$) and the two-dimentional $(\alpha,\phi)$-PESs computed with SA-OO-VQE.

Considering now the non-adiabatic couplings (panel~(c)
of Fig.~\ref{fig:formaldimine_all_90}), we find that it points to a direction perpendicular to the gradients.
This is expected, as the NAC together with the gradient of the energy difference span the branching space defined by $\alpha$ and $\phi$, as discussed in Sec.~\ref{subsubsec:BO}.
The NAC exhibits an asymptotic discontinuity around the conical intersection, caused by the term $(E_1-E_0)^{-1}$ which goes to infinity at this degeneracy point. 
The increase of the NAC amplitude at the avoided crossing or its divergence at the conical intersection is the expected manifestation of its linear dependence on the inverse of the energy difference [see Eq.~(\ref{eq:form_NAC_HF})]. 
This typically results in a break-down of the Born--Oppenheimer approximation, which is consistent with observing radiationless population transfer from the first-excited state back to the ground state. 

This will essentially occur around such crossing geometries along the photochemical reaction path. 
A prototypical example is the photoisomerization process in the retinal chromophore of rhodopsin~\cite{gozem2017theory}. 
In such a situation, quantum dynamics simulations (or their various flavours of quantum-classical approximations) must explicitly account for NAC-terms within the equations of motion for the nuclei evolving within a manifold of coupled electronic states.

\begin{figure*}
\centering
    \includegraphics[width=\textwidth]{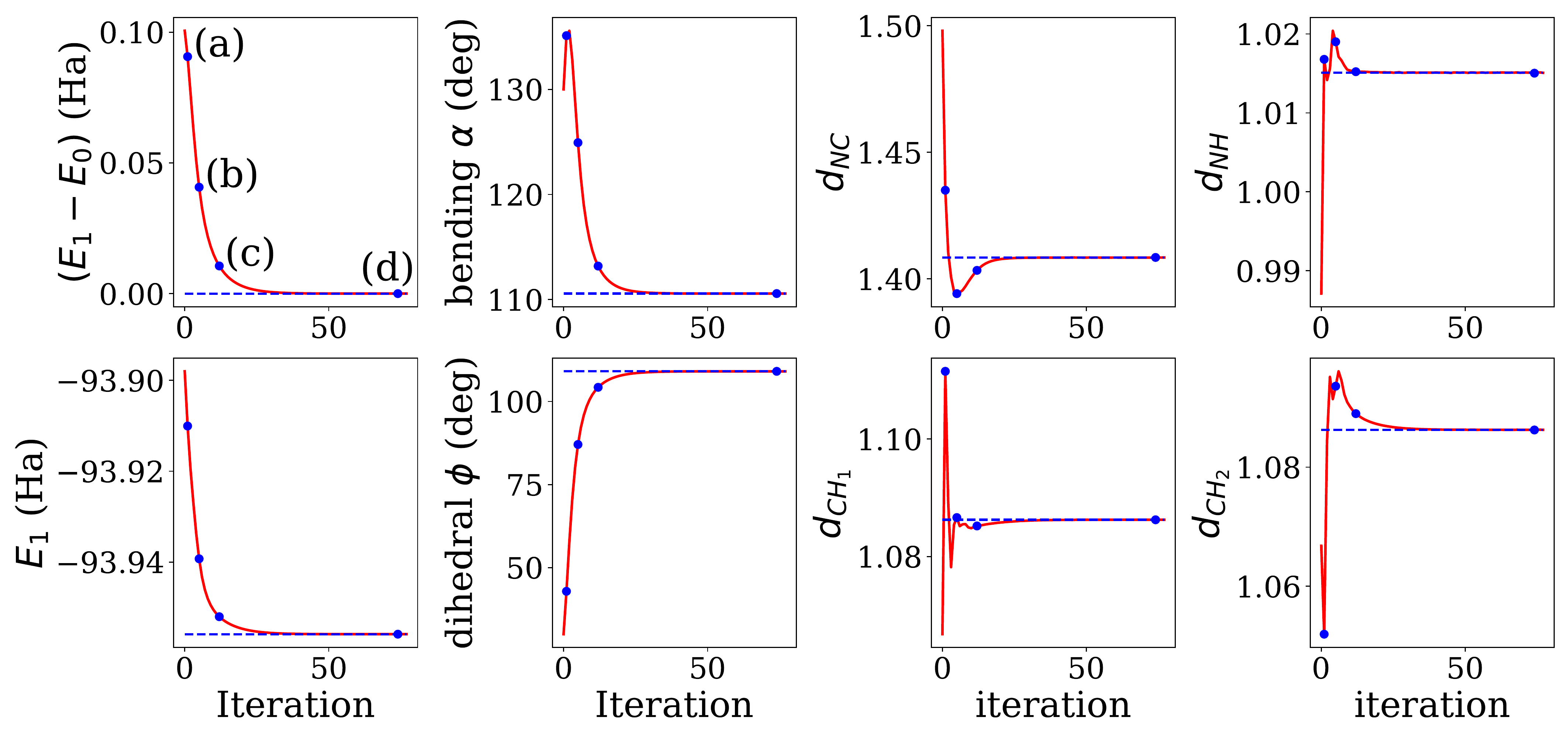}
    \includegraphics[width=\textwidth]{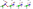}
\caption{\textbf{Geometry optimization to minimal energy conical intersection (MECI)}.
\textbf{Upper panel:} Energy difference, first-excited state energy, bending angle, dihedral angle, and bond lengths of the molecule are plotted against the number of iterations to convergence of the MECI optimization using the SA-OO-VQE algorithm (red lines). 
Dashed blue lines indicate the converged values of the MECI optimization using the SA-CASSCF algorithm.
\textbf{Lower panel:} Four molecular geometries corresponding to four points of the optimization path (blue points on the upper panel). 
The corresponding gradient vectors $-\mathbf{g}$ are shown for each atom.
\label{fig:MECP_optimization}}
\end{figure*}

\subsection{Geometry optimization to locate formaldimine's conical intersection in the $(\alpha,\phi)$ space}\label{subsec:geom_opt}

As an application to the SA-OO-VQE analytical gradients, 
we perform a geometry optimization inside the ($\alpha,\phi$)-plane to find a conical intersection for formaldimine.
To do so, we use a steepest-descent optimization algorithm considering as a cost function the energy difference $\Delta E = E_1^\text{SAOOVQE}-E_0^\text{SAOOVQE}$. 
At each step of the run, we update the molecular geometry in the ($\alpha,\phi$)-plane based on the associated gradient 
\begin{equation}
    \mathbf{g}_{\Delta E } = \dv{ E_1^\text{SAOOVQE}}{ \mathbf{x}} - \dv{ E_0^\text{SAOOVQE}}{ \mathbf{x}}
\end{equation}
which is evaluated using our analytical method described in Sec.~\ref{subsec:gradients_NAC_QC} (with $\mathbf{x} = (\alpha, \phi)$).

An example run of the algorithm is shown in Fig.~\ref{fig:CI_optimization}, where the steepest-descent procedure starts at point~(a) for a molecular configuration $(\alpha,\phi)=(130^\circ,35^\circ)$. 
The path followed during the geometry optimization is driven by the vector $-\mathbf{g}_{\Delta E}$
which is illustrated at every 4 iterations with arrows on the left panel of Fig.~\ref{fig:CI_optimization} (out of a total of 20 iterations required to reach a geometry very near the conical intersection point). The last iteration point is also shown, located exactly on the CI.
Similarly, on the right panel of Fig.~\ref{fig:CI_optimization} the vector $-\mathbf{g}_{\Delta E}$ is represented with red arrows (starting from the Hydrogen atom) for four molecular geometries (a), (b), (c) and (d) obtained on the path of optimization (also noted on the left panel).
From Fig.~\ref{fig:CI_optimization}, we see that the path stays orthogonal to isolines ($\Delta E = c$ with $c$ constant) of the cost function $\Delta E$. 
This feature indicates that the SA-OO-VQE algorithm provides consistent pairs of gradients for the two states, thus leading smoothly to the molecular geometry associated to the conical intersection in the ($\alpha,\phi$)-plane. 
At the end of the geometry optimization,
the conical intersection is found for a geometry $\phi = 90 ^\circ$ and $\alpha = 121.47^\circ$. 
As a comparison, a similar calculation has been realized with the SA-CASSCF method for which a similar path was followed in the ($\alpha,\phi$)-plane, thus leading to an equivalent location of the conical intersection with a negligible difference of the final molecular geometry ($\sim 0.001^\circ$ error for both $\alpha$ and $\phi$).
In practice, note that the non-smooth behavior of the cost function at the conical intersection makes the gradient-descent algorithm hard to converge when approaching this point (as the gradient of the cost function will always have a non-zero component in either direction). 
As the scope of this paper is to provide proof-of-principle calculation, we simply stopped the algorithm after a limited number of iterations. 

\subsection{Conical intersection optimization in the full geometry space}\label{sec:MECI_opt}

The previous section focused on the localization of a conical intersection within the $(\alpha,\phi)$ 2D-subspace
at fixed values of the other internal
coordinates. 
As this subspace is a
good description of the branching space, there is 
a unique point of degeneracy (a conical intersection)
within this plane where degeneracy
is lifted to first order from it. 
As such,
minimising $\Delta E$, and nothing else, is sufficient,
as the complement space is
frozen.

In this last section, a step further is realized to characterize the conical intersection of the formaldimine molecule, by determining the so-called \textit{minimal energy conical intersection} (MECI) of the system. 
In practice, the MECI point corresponds to the optimal geometry of a given system for which the energies of two states get simultaneously degenerated and maximally lowered. 
Therefore, it encodes the most favorable molecular conformation for the realization of non-radiative photochemical processes but also inter-system (i.e. spin-forbidden) crossings~\cite{yarkony1993systematic,yarkony1992theoretical}. As a counterpart, the realization of a geometry optimization to precisely determine the MECI position is usually a pretty involved task. 
The process implies a full relaxation of the internal coordinates of a molecular system, that is driven by the gradients and NAC vectors of each atom. 

Using the estimation of NAC and energy gradients with SA-OO-VQE, we realized a geometry optimization to determine the MECI of formaldimine, using the so-called gradient projection method~\cite{bearpark1994direct,keal2007comparison,sicilia2008new}. 
In this algorithm, we simultaneously minimize $\Delta E^2$ -- to allow for a smooth minimum -- and $E_1$,
using a steepest-descent algorithm where the minimization follows the direction of a composite gradient.
To minimize $E_1$ only in the direction of the seam space, we project out the component of its gradient along the branching space. 
The composite gradient is:
\begin{align}\label{eq:MECP_grad}
    \mathbf{g} = \eta \left( 2 \Delta E \frac{\mathbf{g}_{\Delta E}}{|\mathbf{g}_{\Delta E}|} \right) + (1-\eta)\mathbf{P} \dv{E_1^{\rm SAOOVQE}}{\mathbf{x}},
\end{align}
where $\eta \in [0,1]$ is a constant, balancing the two objectives, and
\begin{align}
    \mathbf{P} = \mathbf{1} - \tilde{\mathbf{g}}_{\Delta E} \tilde{\mathbf{g}}_{\Delta E}^\dagger - \tilde{\mathbf{h}}_{01} \tilde{\mathbf{h}}_{01}^\dagger
\end{align}
is the projection along the seam space. Here, $\mathbf{h}_{IJ}$ is proportional to the CI-term of the NAC, i.e. the first term in Eq.~\eqref{eq:NAC_final}. The tilde indicates orthonormalization of $\mathbf{g}_{\Delta E}$ and $\mathbf{h}_{01}$. 

Results are laid out in Fig.~\ref{fig:MECP_optimization}. We initialize the algorithm with $\eta=0.25$ and consider convergence thresholds of $10^{-13}$~Ha$^2$ and $10^{-6}$~Ha for $\Delta E^2$ and $E_1$, respectively. 
As in the previous section, the starting geometry is $(\alpha,\phi)=(130^\circ,35^\circ)$, but now all the $3N_{\rm atom}$ nuclear coordinates 
move in the direction of their respective gradient defined by Eq.~\eqref{eq:MECP_grad}. 
Our algorithm converges after 78 iterations, where the bending and dihedral angles are $(\alpha,\phi)=(110.6^\circ,109.1^\circ)$. 
Note that a dihedral angle of $109.1^\circ$ here does not break the $\sigma_v$ mirror symmetry of the molecule, as the hydrogens attached to the carbon bend slightly backwards.

As a reference, in Fig.~\ref{fig:MECP_optimization} dashed lines are used to represent the final results obtained for a similar MECI optimization realized with the SA-CASSCF method (analytical SA-CASSCF gradients and NAC being evaluated with OpenMolcas to drive the optimization). 
As readily seen in the different plots, the MECI optimization based of SA-OO-VQE (red lines) converges to the same geometry and energies as in SA-CASSCF,
thus showing again the accuracy of our estimations of the NAC and energy gradients within the SA-OO-VQE algorithm. 

\section{Conclusions and perspectives}\label{sec:conclu}

In this paper, we introduce several tools to improve our original SA-OO-VQE algorithm~\cite{Yalouz_2021}.
The first improvement consists in introducing a flexible and efficient way to resolve the SA-OO-VQE electronic states.
The method, based on equi-ensemble properties, takes advantage of the invariance of the state-averaged energy under rotation of the states involved in the ensemble. 
Using a simple rotation of the input states (implemented by a short-depth circuit), we show that one can postpone the resolution of the electronic states to the very end of the full SA-OO-VQE scheme,
thus avoiding many unnecessary manipulations and quantum measures during intermediate steps of the algorithm. 
The second improvement is the development of theoretical methods to extract analytical derivatives within the SA-OO-VQE algorithm. 
These derivatives --
the nuclear energy gradients and non-adiabatic couplings --
are fundamental for the study of molecular systems,
and can be determined using
Lagrangian methods that are intimately linked to the so-called coupled-perturbed theory.
The accuracy of our derivations
is checked against reference results based on SA-CASSCF calculations, for which we obtain very good agreement.
Finally, we illustrate the use of these derivatives in practical calculations
by performing the geometry optimization towards the conical intersection of the formaldimine molecule.
The localization of the spectral degeneracy matches perfectly the predictions from the SA-CASSCF
method.

The definition of these new tools opens the way to several new developments. 
One aspect that we already briefly touched upon above, is the definition of the intermediate diabatic and final adiabatic bases. 
In our implementation of the SA-OO-VQE algorithm, the procedure starts from reference guess states and has no reason to produce ``excessive'' transformations,
thus making the least-transformed subspace a good candidate for being a quasidiabatic representation (see Ref.~\citenum{cederbaum1989block}). 
At the moment, we have been observing such a property, and we have good incentive but no formal proof. 
Further work is under way to show that SA-OO-VQE before its final diagonalization (or state-resolution) could indeed be an efficient avenue for the \emph{ab initio} production of relevant quasidiabatic states.
Such results would be important as it would facilitate use in molecular quantum dynamics applications. 
With an appropriate definition of such states and construction of a quantum-classical interface, these tools can benefit from classical implementations~\cite{Richter2011JCTC} of algorithms to perform various forms of molecular dynamics~\cite{Tully,Andrade.Alonso.2009bjl}, thus giving quantum co-processing a firm place in the toolbox of quantum chemical simulations. 
A related aspect is to consider different kinds of surface couplings, like those provided by the spin-orbit operator. 
This is of interest for a range of applications, e.g. the rate of inter-system crossing that is key to technological applications, as the construction of more efficient blue light emitting diodes. 
Here, we note that it is possible to work with a different set of trial states than the Hartree--Fock and the singlet excited model wave functions chosen in the current work. 
The latter can be replaced by a triplet excited wave function, while the former can also be a non-ground state determinant (or other simple wave function) if we are interested in excited-state couplings.

\section*{Acknowledgments}  
SY acknowledges support from the Netherlands Organization for Scientific Research (NWO/OCW), and the Interdisciplinary Thematic Institute ITI-CSC
via the IdEx Unistra (ANR-10-IDEX-0002) within the program Investissement d’Avenir. EK acknowledges support from Shell Global Solutions BV. 

\section*{Note added}

As we were finalizing the writing process of this manuscript, a similar work appeared on the arXiv~\cite{arimitsu2021analytic}.
While they show how to estimate analytical gradients in several excited-state extensions of VQE
(which we only do for SA-OO-VQE),
our work adds several
features to theirs, such as the analytical estimation of non-adiabatic couplings
and a discussion on the capture of diabatic versus non-adiabatic states within SA-OO-VQE.
This paper and Ref.~\citenum{arimitsu2021analytic}
are therefore complementary, and they both pave the way towards excited-state quantum dynamics with excited-state VQE extensions.

\newcommand{\Aa}[0]{Aa}
%


\appendix

\section{Circuit gradient $\mathbf{G}^{\rm C}$ and Hessian $\mathbf{H}^{\rm CC}$}\label{app:GC_HCC}

Let us consider the estimation of the expectation value $M$ of a generic operator $\hat{\mathcal{M}}$ (in our case the electronic structure Hamiltonian $\hat{\mathcal{H}}$)
with respect to a state $\ket{\Psi(\bmtheta)} = \hat{U}(\bmtheta) \ket{\Phi_0}$,
\begin{equation}
    M(\bmtheta) =  {\rm Tr}\left[\hat{\mathcal{M}}\rho(\bmtheta) \right] =  \bra{\Psi(\bmtheta)}\hat{\mathcal{M}}\ket{\Psi(\bmtheta)},
\end{equation}
where $\rho(\bmtheta) = \ket{\Psi(\bmtheta)}\bra{\Psi(\bmtheta)}$ is the density matrix operator
and ${\rm Tr}\left[\cdot\right]$ the trace operation.
These matrix elements encode the first order (gradient $\mathbf{G}^{\rm C}$) and second order (Hessian $\mathbf{H}^{\rm CC}$) derivatives with respect to the ansatze parameters $\bmtheta$ read as follows
\begin{equation}
    G^{\rm C}_{i} = \frac{\partial  {M}(\bmtheta)}{\partial \theta_i}  \quad \text{ and } \quad H^{\rm CC}_{ij} = \frac{\partial^2  {M}(\bmtheta)}{\partial \theta_i \partial \theta_j},
\end{equation}
and can be evaluated
with the parameter-shift rule~\cite{li2017efficient,mitarai2018quantum,schuld2019evaluating,mari2020estimating,meyer2021gradients,hubregtsen2021single}.

As a starting point, let us consider the following unitary:
\begin{equation}
\begin{split}
    U(\bmtheta) &= U_1(\theta_1) \times U_2(\theta_2)\\
    &= (e^{-i\frac{\theta_1}{2}P_1}e^{-i\frac{\theta_1}{2}\tilde{P}_1}) \times (e^{-i\frac{\theta_2}{2}P_2}e^{-i\frac{\theta_2}{2}\tilde{P}_2})
\end{split}
\end{equation}
where $[P_1,\tilde{P}_1] = [P_2,\tilde{P}_2] = 0$ with the tilde notation denoting a different Pauli string with the same associated parameters, as this is usually the case in the fermionic-UCC ansatz.
We have $P_j = P_j^\dagger$ and
\begin{eqnarray}
\dfrac{\partial U_j(\theta_j)}{\partial \theta_j}
= - \dfrac{i}{2} (P_j + \tilde{P}_j) U_j(\theta_j)
\end{eqnarray}
and
\begin{eqnarray}
\left(\dfrac{\partial U_j(\theta_j)}{\partial \theta_j}\right)^\dagger
= - \dfrac{\partial U_j(\theta_j)}{\partial \theta_j}.
\end{eqnarray}
We want to estimate the gradient elements 
\begin{eqnarray}
G^{\rm C}_{i}
= \dfrac{\partial}{\partial \theta_i}  {\rm Tr}\left[\hat{\mathcal{M}}\rho(\bmtheta) \right].
\end{eqnarray}
By taking the derivative with respect to $\theta_2$, we get
\begin{eqnarray}\label{eq:A}
G^{\rm C}_{2}  &=&  {\rm Tr}\left[ \dfrac{\partial}{\partial \theta_2}  \hat{\mathcal{M}} \rho(\bmtheta) \right]\nonumber\\
 &=& \hat{\mathcal{M}} U_1(\theta_1) \left[ \dfrac{\partial U_2(\theta_2)}{\partial \theta_2} \rho U_2^\dagger(\theta_2) \right. \nonumber\\
&&\left. + U_2(\theta_2) \rho \left(\dfrac{\partial U_2(\theta_2)}{\partial \theta_2}\right)^\dagger \right] U_1^\dagger(\theta_1)\nonumber \\
&=& - \dfrac{i}{2} \hat{\mathcal{M}} U_1(\theta_1) \left[(P_2 + \tilde{P}_2) U_2(\theta_2) \rho U_2^\dagger(\theta_2) \right. \nonumber \\
&& \left. - U_2(\theta_2) \rho (P_2 + \tilde{P}_2) U_2^\dagger(\theta_2) \right] U_1^\dagger(\theta_1)\nonumber\\
&=& - \dfrac{i}{2} \hat{\mathcal{M}} U_1(\theta_1)U_2(\theta_2) \left[(P_2 + \tilde{P}_2), \rho \right] U_2^\dagger(\theta_2) U_1^\dagger(\theta_1)\nonumber \\
\end{eqnarray}
where we used the property $[P_i, U_i(\theta_i)] = 0$
and the notation $\rho = \ket{\Phi_0}\bra{\Phi_0}$.
We then use the property of commutator for an arbitrary operator [see Eq.~(2) in Ref.~\citenum{mitarai2018quantum}, where $U_j(\theta_j) = \exp(- i \theta_j P_j/2)$],
\begin{eqnarray}\label{eq:unknown}
[P_j,\rho] = i \left[U_j\left(\dfrac{\pi}{2}\right) \rho U_j^\dagger\left(\dfrac{\pi}{2}\right) - U_j\left(-\dfrac{\pi}{2}\right)
\rho U_j^\dagger\left(-\dfrac{\pi}{2}\right)\right],\nonumber\\
\end{eqnarray}
which can be easily demonstrated by considering the property of exponential of Pauli strings
\begin{equation}
    e^{i\frac{\theta_j}{2} P_j} = \cos\left(\frac{\theta_j}{2}\right) \mathbf{1} - i \sin\left(\frac{\theta_j}{2}\right) P_j.
\end{equation}
One then separates $(P_j + \tilde{P}_j)$ as follows:
\begin{eqnarray}
[(P_j + \tilde{P}_j), \rho] = [P_j, \rho] + [\tilde{P}_j,\rho]
\end{eqnarray}
such that, by inserting Eq.~(\ref{eq:unknown}) into Eq.~(\ref{eq:A}), the gradient reads
\begin{eqnarray}
G^{\rm C}_{2}  &=& \dfrac{1}{2} \hat{\mathcal{M}} U_1(\theta_1) \left(
e^{-\frac{i}{2}\left(\theta_2 + \frac{\pi}{2}\right)P_2}
e^{-\frac{i}{2}\theta_2\tilde{P}_2}
\rho e^{\frac{i}{2}\theta_2\tilde{P}_2}
e^{\frac{i}{2}\left(\theta_2 + \frac{\pi}{2}\right)P_2}\right.\nonumber \\
&&
\left.
- e^{-\frac{i}{2}\left(\theta_2 - \frac{\pi}{2}\right)P_2}
e^{-\frac{i}{2}\theta_2\tilde{P}_2}
\rho e^{\frac{i}{2}\theta_2\tilde{P}_2}
e^{\frac{i}{2}\left(\theta_2 - \frac{\pi}{2}\right)P_2}\right.\nonumber \\
&&
\left.
+ e^{-\frac{i}{2}\theta_2 P_2}
e^{-\frac{i}{2}\left(\theta_2 + \frac{\pi}{2}\right)\tilde{P}_2}
\rho e^{\frac{i}{2}\left( \theta_2  + \frac{\pi}{2}\right) \tilde{P}_2}
e^{\frac{i}{2}\theta_2P_2}\right.\nonumber \\
&&
\left.
- e^{-\frac{i}{2}\theta_2 P_2}
e^{-\frac{i}{2}\left(\theta_2 - \frac{\pi}{2}\right)\tilde{P}_2}
\rho e^{\frac{i}{2}\left( \theta_2  - \frac{\pi}{2}\right) \tilde{P}_2}
e^{\frac{i}{2}\theta_2P_2}
\right) U_1^\dagger(\theta_1) \nonumber \\
\end{eqnarray}
which, by taking the trace of it, leads to the final expression 
\begin{eqnarray}\label{eq:GC2}
G^{\rm C}_{2}  =
\dfrac{1}{2}\left( \expval{\hat{\mathcal{M}}}_{\theta_2^+} - \expval{\hat{\mathcal{M}}}_{\theta_2^-}
+ \expval{\hat{\mathcal{M}}}_{\tilde{\theta}_2^+} - \expval{\hat{\mathcal{M}}}_{\tilde{\theta}_2^-}\right)\nonumber \\
\end{eqnarray}
where the notation $\expval{\hat{\mathcal{M}}}_{\theta_j^\pm}$ refers to the expectation value of the operator $\hat{\mathcal{M}}$ when $\theta_j$ has been shifted by $\pm \pi/2$ in front of $P_j$ (and in front of $\tilde{P}_j$ for $\expval{\hat{\mathcal{M}}}_{\tilde{\theta}_j^\pm}$).

Generalizing Eq.~(\ref{eq:GC2}) to any
$n$-fold fermionic excitation generator
$\mathcal{G}_j$ associated to the parameter $\theta_j$ leads to the parameter-shift rule:
\begin{eqnarray}\label{eq:parameter_shift_rule_gradient}
G^{\rm C}_{j} 
= \dfrac{1}{2}\sum_{n}^{\forall P_n \in \mathcal{G}_j} \left( \expval{\hat{\mathcal{M}}}_{\theta_{j_n}^+} - \expval{\hat{\mathcal{M}}}_{\theta_{j_n}^-} \right)
\end{eqnarray}
where $\theta_{j_n}$ refers to the parameter associated to the Pauli string $P_n$ coming from the fermionic generator $\mathcal{G}_j$.
Note that the parameter-shift rule
only applies to generators $\mathcal{G}$ that have at most two distinct eigenvalues~\cite{schuld2019evaluating}, which is always the case for any Pauli string but not a linear combination of them.
However, each parameter $\theta_j$ is associated to two Pauli strings for a single-excitation fermionic operator, to eight Pauli strings for a double-excitation fermionic operator, and to $2^{2n-1}$ Pauli strings for a $n$-fold fermionic excitation operator~\cite{romero2018strategies,kottmann2020feasible}, such that
\begin{eqnarray}
e^{{\rm i}\frac{\theta}{2}\mathcal{G}} \rightarrow \prod_{x=1}^{2^{2n-1}} e^{{\rm i}\frac{\theta}{2}P_x}.
\end{eqnarray}
The above formula is actually an equality because Pauli strings resulting from a same $n$-fold fermionic excitation operator actually commute with each other~\cite{romero2018strategies}.
Although the fermionic generator $\mathcal{G}$ usually doesn't have two distinct eigenvalues
but three~\cite{kottmann2020feasible},
they can be decomposed into
generators that have only two distinct eigenvalues
(for instance, Pauli strings $P_x$ with eigenvalues $\pm$1) and the gradient can be directly calculated by the product rule and the parameter-shift rule~\cite{crooks2019gradients,kottmann2020feasible}, necessitating $2^{2n}$ expectation values.
So even for a UCCD ansatz, we would need around $2^4 = 16$ expectation values for a single gradient calculation (to be multiplied by the number of parameters).

Turning to the Hessian estimation, one can derive Eq.~(\ref{eq:parameter_shift_rule_gradient}) with respect to another parameter $\theta_k$,
\begin{eqnarray}
H^{\rm CC}_{kj}
= \dfrac{1}{2} \sum_{n}^{\forall P_n \in \mathcal{G}_j} \left( \dfrac{\partial }{\partial \theta_k} \expval{\hat{\mathcal{M}}}_{\theta_{j_n}^+}- \dfrac{\partial }{\partial \theta_k} \expval{\hat{\mathcal{M}}}_{\theta_{j_n}^-} \right) \nonumber \\
\end{eqnarray}
and use the parameter-shift rule again,
thus leading to~\cite{mari2020estimating}
\begin{equation}\label{eq:Hess_ii}
\begin{split}
H^{\rm CC}_{kj} = \dfrac{1}{4} \sum_{n}^{\forall P_n \in \mathcal{G}_j}   \sum_{m}^{\forall P_m \in \mathcal{G}_k} 
\Big( \expval{\hat{\mathcal{M}}}_{\theta_{j_n}^+\theta_{k_m}^+} - \expval{\hat{\mathcal{M}}}_{\theta_{j_n}^+\theta_{k_m}^-} \\
-\expval{\hat{\mathcal{M}}}_{\theta_{j_n}^-\theta_{k_m}^+} + \expval{\hat{\mathcal{M}}}_{\theta_{j_n}^-\theta_{k_m}^-}
 \Big). \\
\end{split}
\end{equation}
According to Eq.~(\ref{eq:Hess_ii}), a single element of the Hessian will require the estimation of $2^{4n}$ expectation values for a $n$-fold fermionic excitation operator.

Different strategies have recently been developed to reduce the number of expectation values required
to evaluate a ansatze-parameter gradient of a fermionic generator $\mathcal{G}$ with more than two distinct eigenvalues.
One can consider an additional ancilla qubit and decompose the derivative into a linear combination of unitaries~\cite{schuld2019evaluating},
use stochastic strategies~\cite{banchi2020measuring,wierichs2021general}
or different generator decomposition techniques~\cite{kottmann2020feasible,izmaylov2021analytic}.

\section{ Orbital gradient $\mathbf{G}^{\rm O}$ and Hessian $\mathbf{H}^{\rm OO}$ } 

In this appendix, we show how the orbital gradient and Hessian can be estimated from the one- and two-particle reduced density matrices (1-RDM and 2-RDM) that
are measured out of the quantum circuit.
For simplicity, let us focus on single wavefunction $| \Psi_I (\boldsymbol{\theta})\rangle $ for which we want to optimize the orbitals.
The generalization to a weighted-ensemble of state (as in SA-OO-VQE) is straightforward, as one just has to replace the state-specific 1- and 2-RDMs by the
state-averaged 1- and 2-RDMs.
The parametrized energy of the state reads
\begin{equation}
\begin{split}
    E_I(\boldsymbol{\kappa},\boldsymbol{\theta}) &= \bra{\Psi_I(\boldsymbol{\theta})}   \hat{U}_\text{O}(\boldsymbol{\kappa})^\dagger  \hat{\mathcal{H}} \hat{U}_\text{O}(\boldsymbol{\kappa})  \ket{\Psi_I(\boldsymbol{\theta})},
\end{split}
\end{equation}
where the orbital rotation operator is defined such that
\begin{equation}
    \hat{U}_\text{O}(\boldsymbol{\kappa}) = e^{-\hat{\kappa}} \text{,  with }  \hat{\kappa} = \sum_{p>q}  \kappa_{pq} \hat{E}_{pq}^-.
\end{equation}
with $\hat{E}_{pq}^- = \hat{E}_{pq} - \hat{E}_{qp}$. Expanding to second order in $\boldsymbol{\kappa}$ the operator $\hat{U}_\text{O}(\boldsymbol{\kappa})^\dagger \hat{\mathcal{H}} \hat{U}_\text{O}(\boldsymbol{\kappa}) $ leads to
\begin{equation} 
    E_I(\boldsymbol{\kappa},\boldsymbol{\theta}) \simeq \bra{\Psi_I(\boldsymbol{\theta})} \Big( \hat{\mathcal{H}} +  [\hat{\kappa}, \hat{\mathcal{H}}]   + \frac{1}{2}   \big[ \hat{\kappa},[\hat{\kappa} ,\hat{\mathcal{H}} ]\big]\Big) \ket{\Psi_I(\boldsymbol{\theta})},
\label{dev1}
\end{equation}
which, when compared to the second-order Taylor expansion of $E_I(\boldsymbol{\kappa})$ with respect to the $\boldsymbol{\kappa}$ parameters,
\begin{equation}
\begin{split}
    E_I( \boldsymbol{\kappa},\boldsymbol{\theta}) &\simeq E_I(0,\boldsymbol{\theta}) +\boldsymbol{\kappa} ^\dagger\mathbf{G}^{\text{O},I} + \frac{1}{2} \boldsymbol{\kappa}^\dagger \mathbf{H}^{\text{OO},I}\boldsymbol{\kappa},
\end{split}
\label{dev2}
\end{equation}
allows to identify
the MO-gradient and MO-Hessian elements, defined as follows:
\begin{eqnarray}
{G}_{pq}^{\text{O},I} = \dfrac{\partial E_I}{\partial \kappa_{pq} } = \bra{\Psi_I (\boldsymbol{\theta})} [\hat{E}_{pq}^-,\hat{\mathcal{H}}] \ket{\Psi_I(\boldsymbol{\theta})}
\label{eq:OO_gradient}
\end{eqnarray}
for the gradient, and
\begin{eqnarray}
{H}_{pq,rs}^{\text{OO},I} &=& \dfrac{\partial^2 E_I}{\partial \kappa_{pq}\partial\kappa_{rs} } \nonumber \\
&=&  \dfrac{1}{2}\left(1+\mathcal{S}_{(pq)}^{(rs)}\right)\bra{ \Psi_I (\boldsymbol{\theta})} \big[\hat{E}_{pq}^-,  [\hat{E}_{rs}^-,\hat{\mathcal{H}}]\big] \ket{\Psi_I (\boldsymbol{\theta})} \nonumber \\
    \label{eq:OO_hessian}
\end{eqnarray}
for the Hessian,
where $\mathcal{S}_{(pq)}^{(rs)}$ is an operator that permutes the two couples of indices $(pq)$ and $(rs)$. 
In practice, one can derive an analytic form of the orbital gradient and Hessian based on the 1 and 2-RDMs and the electronic integrals. 
As this derivation is fastidious (and already available in the literature~\cite{helgaker2014molecular}), we only introduce the final equations
required for the implementation.
Starting with the orbital gradient  $\mathbf{G}^{\text{O},I}$, the elements of the associated matrix read
\begin{equation}
{G}_{pq}^{\text{O},I}=  2({F}_{pq}^I- {F}_{qp}^I),
\end{equation}
where the elements of the generalized Fock matrix $\mathbf{F}^I$ (associated to the state $\ket{\Psi_I}$) read
\begin{equation}\label{eq:Fock_matrix}
{F}_{pq}^I = \sum_{t} \gamma_{pt}^I \ h_{qt} +\sum_{t,u,v} \Gamma_{ptuv}^I \  g_{qtuv}.
\end{equation}
In practice, building the full matrix can be very expensive. 
However and as described in Ref.~\citenum{helgaker2014molecular}, considering an active space partitioning does reduce this complexity considerably.
In this partitioning, the Fock matrix is fragmented into three contributions:
\begin{eqnarray}
 {F}_{iq}^I  &=& 2( {F}_{qi}^{\text{frozen},I} + {F}_{qi}^{\text{active},I} ),\\
{F}_{vq}^I   &=& \sum_w^{\text{active}}  {F}_{qw}^{\text{frozen},I} \gamma_{vw}^I + \sum_{w,x,y}^{{\text{active}}}  \Gamma_{v wxy}^I g_{q wxy} ,
\end{eqnarray}
and
\begin{eqnarray}
{F}_{aq}^I   &= 0 ,
\end{eqnarray}
where $i,v$ and $a$ refer to frozen occupied, active and virtual MOs, respectively,
and
$\mathbf{F}^{\text{frozen},I}$ and $\mathbf{F}^{\text{active},I}$ are the so-called frozen and active Fock matrices that read
\begin{eqnarray}
{F}^{\text{frozen},I}_{pq} &= h_{pq} + \sum_i^{\text{frozen}}  (2g_{pqii}-g_{piiq})
\end{eqnarray}
and
\begin{eqnarray}
{F}^{\text{active},I}_{pq}  &= \sum_{w,x}^{\text{active}}  \gamma_{wx}^I(g_{pqwx}-\frac{1}{2}g_{pxwq}).
\end{eqnarray}
Turning to the orbital Hessian $\mathbf{H}^{\text{OO},I}$, the elements of the associated matrix read
\begin{equation}
\begin{split}
     {H}_{pq,rs}^{\text{OO},I}  &= (1-\mathcal{S}_{pq})(1-\mathcal{S}_{rs}) \Big\lbrace  ({F}_{ps}^I+{F}_{sp}^I)\delta_{qr} - 2h_{ps}\gamma_{qr}^I \\
        &+2\sum_{t,u}\big(g_{purv}(\Gamma_{qusv}^I+\Gamma_{quvs}^I) + g_{pruv}\Gamma_{qsuv}^I\big)\Big\rbrace
\end{split}
\end{equation}
where $\mathcal{S}_{pq}$ ($\mathcal{S}_{rs}$) is an operator permuting the indices $p$ and $q$ ($r$ and $s$).

Note that, within the active space approximation,
only the 1- and 2-RDM elements from the active space
have to be measured on the quantum computer.
Every other non-zero terms of the RDMs read 
\begin{eqnarray}
    \gamma_{ij}^I =\gamma_{ji}^I  &=&  2 \delta_{ij} \\
    \Gamma_{ijkl}^I &=& 4\delta_{ij}\delta_{kl}-2\delta_{il}\delta_{jk}\\
    \Gamma_{ijwx}^I = \Gamma_{wxij}^I &=& 2 \gamma_{wx}^I\delta_{ij} \\
    \Gamma_{iwxj}^I = \Gamma_{xjiw}^I &=& - \gamma_{wx}^I\delta_{ij} 
\end{eqnarray} 
where $i,j,k,l$ and $w,x$ denote frozen and active MO indices, respectively.

\section{Circuit-orbital hessian $\mathbf{H}^{\rm CO}$}

In practice, one can estimate the off-diagonal blocks of the Hessian matrix $\mathbf{H}^\text{CO}$ by repeated measurements of the quantum circuit.
From the definition of the molecular orbital gradient in Eq.~(\ref{eq:OO_gradient}),
one obtains
\begin{equation}
\begin{split}
  {H}^{\text{CO},I}_{j,pq} &= \frac{\partial}{\partial \theta_j } \frac{\partial E_I}{ \partial\kappa_{pq} } \\
   &= \frac{\partial }{\partial \theta_j }  \bra{\Psi_I(\boldsymbol{\theta})} [\hat{E}_{pq}^-, \hat{\mathcal{H}}] \ket{\Psi_I(\boldsymbol{\theta})}\\
&= \frac{\partial }{\partial \theta_j } {\rm Tr}\left[ \hat{\mathcal{M}} \hat{\rho}_I(\boldsymbol{\theta}) \right],
\end{split}
\end{equation}
where $\hat{\rho}_I(\boldsymbol{\theta}) = \ket{\Psi_I(\boldsymbol{\theta})}\bra{\Psi_I(\boldsymbol{\theta})}$ and $\hat{\mathcal{M}}=[\hat{E}_{pq}^-, \hat{\mathcal{H}}]$.
Using the parameter-shift rule on the operator $\hat{\mathcal{M}}$ (see Appendix \ref{app:GC_HCC}), we obtain the elements of $\mathbf{H}^{\text{CO},I}$ as
\begin{eqnarray}\label{eq:HCO_I}
 {H}^{\text{CO},I}_{j,pq} =
\dfrac{1}{2}\sum_{n}^{\forall P_n \in \mathcal{G}_j} \left( \expval{\hat{\mathcal{M}}}_{\theta_{j_n}^+} - \expval{\hat{\mathcal{M}}}_{\theta_{j_n}^-} \right).
\end{eqnarray}
According to Eq.~(\ref{eq:HCO_I}), one 
can measure the expectation values of the new operator  $\hat{\mathcal{M}}=[\hat{E}_{pq}^-, \hat{\mathcal{H}}]$,
with the appropriate shift-in-parameter defined from the parameter-shift rule.
In practice, Eq.~(\ref{eq:HCO_I}) can be rewritten in terms of generalized Fock matrices,
\begin{equation} 
     {H}^{\text{CO},I}_{j,pq} = \sum_{n}^{\forall P_n \in G_j}  \Big( F_{\theta_{j_n}^+,pq}^I-F_{\theta_{j_n}^+,qp}^I  - F_{\theta_{j_n}^-,pq}^I + F_{\theta_{j_n}^-,qp}^I \Big),
\end{equation}
which elements can be computed according to 
Eq.~(\ref{eq:Fock_matrix}), with the 1- and 2-RDMs
associated to the $\theta_{j_n}^\pm$-shifted state.
The elements of the opposite off-diagonal Hessian block are simply built based on the symmetry $\mathbf{H}^{\rm OC} =  (\mathbf{H}^\text{CO})^{T}$.

\section{Nuclear derivative of the electronic Hamiltonian operator}\label{app:derivative_hamiltonian}

In the coupled-perturbed equations, one needs the derivative of the Hamiltonian operator with respect to a nuclear coordinate~\cite{staalring2001analytical,helgaker1984second,helgaker1988analytical,simons1984higher} which is defined by
\begin{equation}\label{eq:dHdx}
\begin{split}
       \frac{\partial \hat{\mathcal{H}}}{\partial x} &= \sum_{p,q}  \pdv{h_{pq}}{x} \hat{E}_{pq}  +\frac{1}{2} \sum_{p,q,r,s}  \pdv{g_{pqrs}}{x} \hat{e}_{pqrs} + \pdv{E_{nuc}}{x} ,
       \end{split}
\end{equation}
where the derivative of the electronic integrals are  
\begin{align}  
       \pdv{h_{pq}}{x}  &=    h_{pq}^{(x)} - \frac{1}{2} \left\lbrace   S^{(x)} ,h \right\rbrace_{pq}    \\
      \pdv{g_{pqrs}}{x} &=    g_{pqrs}^{(x)} -\frac{1}{2}\left\lbrace S^{(x)} ,g \right\rbrace_{pqrs}   
\end{align}
where we retrieve `explicit' and `response' terms with respect to a nuclear coordinate. The explicit terms are the ones super-scripted with ${}^{(x)}$ indicating a differentiation of the primitive atomic orbitals (MOs coefficients remaining constant). They are defined such as 
\begin{align}
  S_{pq}^{(x)}  &= \sum_{\mu,\nu}^\text{AOs} C_{\mu p } C_{\nu q } \frac{\partial S_{\mu\nu} }{\partial x} \\
  h_{pq}^{(x)}  &= \sum_{\mu,\nu}^\text{AOs} C_{\mu p } C_{\nu q } \frac{\partial h_{\mu\nu} }{\partial x}\\
g_{pqrs}^{(x)}  &= \sum_{\mu,\nu,\delta, \gamma}^\text{AOs} C_{\mu p } C_{\nu q } C_{\delta p } C_{\gamma q } \frac{\partial g_{\mu\nu\delta\gamma} }{\partial x}
\end{align}
where $\mathbf{C}$ is the MO coefficient matrix encoding the optimal orbitals that minimize the state-averaged energy. The `response' terms in curly brackets are defined as
\begin{align}
\left\lbrace   S^{(x)} ,h \right\rbrace_{pq}   = \sum_o ( &S^{(x)}_{po} h_{oq}   + S^{(x)}_{qo}  h_{po} )\\
\left\lbrace   S^{(x)} ,g \right\rbrace_{pqrs} = \sum_o \Big(  &S^{(x)}_{po} g_{oqrs} + S^{(x)}_{qo}  g_{pors}   \\
  + &S^{(x)}_{ro}  g_{pqos} +   S^{(x)}_{so}  g_{pqro}  \Big). 
\end{align}
The last term present on the right of Eq.~(\ref{eq:dHdx}) is the nuclear derivative of the nuclear repulsion energy which is pretty straightforward to compute in practice.
 
\section{Analytical derivation of non-adiabatic couplings for SA-OO-VQE}\label{app:NAC_deriv}

In this section, we introduce the steps to derive the analytical form of Eqs.~(\ref{eq:Lagrange_mult_matrix_NAC}) and (\ref{eq:NAC_final}) which define the NAC between two states $\ket{\Psi_I}$ and $\ket{\Psi_J}$. 
Following Ref.~\citenum{lengsfield1992nonadiabatic}, one splits the complete derivative in the NAC into two contributions,
\begin{align}
    D_{IJ} = \bra{\Psi_I} \ket{\dfrac{d}{dx} \Psi_J} = \bra{\Psi_I} \ket{\pdv{}{x} \Psi_J} + D^{\rm CSF}_{IJ}.
\end{align}
The first term represents the so-called CI term, and the second one the CSF term (see Ref.~\citenum{yarkony1995modern})
{that does not appear in the exact theory [see Eq.~(\ref{eq:def_NAC})].}
The CSF term is readily computed as:
\begin{align}
    D^{\rm CSF}_{IJ} = - &\dfrac{1}{2}\sum_{pq}\gamma_{pq}^{IJ} \big( (\partial_x p|q) - (q | \partial_x p) \big).
\end{align}
The CI term, however, is more involved.
To evaluate this term, we will make use of the off-diagonal Hellmann--Feynman theorem:
\begin{align}\label{eq:Hellmann-Feynman_offdiag}
    \bra{\Psi_I} \ket{\pdv{}{x} \Psi_J} = \Delta E^{-1} \mel{\Psi_I}{\dfrac{\partial \hat{\mathcal{H}}  }{\partial x}}{\Psi_J}.
\end{align}
Eq.~(\ref{eq:Hellmann-Feynman_offdiag}) is valid if two conditions are met. 
The first one is that the SA-OO-VQE states $\ket{\Psi_I}$ and $\ket{\Psi_J}$ are good approximations of the exact eigenstates of $\hat{\mathcal{H}}$ (to some negligible errors, which is verified numerically in our work).
Second, the NAC has to be variational with respect to the orbital rotation parameters $\bmkappa$, the ansatze parameters $\bmtheta$ and the final rotation $\varphi$ implemented for the state resolution. 
While the
SA-OO-VQE states do not satisfy this condition, one can
still make the NAC variational with respect to these parameters by introducing the following Lagrangian:
\begin{equation}\label{eq:Lgrangian_NAC}
\begin{split}
    L_{IJ} &= \overline{S}_{IJ}  \\
           &+ \frac{1}{\overline{\Delta E}} \Bigg( \sum_{p,q} \overline{\kappa}_{pq}^{IJ} \dfrac{\partial E_\text{SA}}{\partial \kappa_{pq}} +  \sum_{n} \overline{\theta}_{n}^{IJ} \dfrac{\partial E_\text{SA}}{\partial \theta_n}
    + \overline{\varphi}^{IJ}\dfrac{\partial \Delta E}{\partial \varphi} \Bigg),
\end{split} 
\end{equation}
where $\overline{S}_{IJ} = \bra{\overline{\Psi_I}}\ket{\Psi_J}$ is the overlap between the two states, and the leftmost state is kept constant (as denoted by an overbar) because we \textit{only} want to take derivative of the right state in the NAC. 
Compared to the gradient Lagrangian [Eq.~(\ref{eq:CP_SAOOVQE_energy_grad})], note the presence of the convergence condition $ \partial \Delta E/\partial \varphi = 0$ encapsulating the effect of the final state resolution (with $\Delta E \equiv  E_J - E_I$). 
In practice, this condition holds as the final rotation consists in minimizing a given state energy (which is equivalent to maximizing the difference between both individual-state energies). 
This convergence condition was not needed in the analytical gradient Lagrangian in Eq.~(\ref{eq:CP_SAOOVQE_energy_grad}) as, in contrast to the NAC, the individual-state energies are already variational with respect to $\varphi$. 
The factor $ \overline{\Delta E}$ in Eq.~(\ref{eq:Lgrangian_NAC}) was introduced for convenience (with the overbar meaning that the energy difference is kept constant).

Now, one has to find the Lagrangian multipliers
in Eq.~(\ref{eq:Lgrangian_NAC}) such that the Lagrangian is
fully variational with respect to $\bmkappa$, $\bmtheta$ and $\varphi$,
\begin{eqnarray}\label{eq:NAC_stationary}
\pdv{L_{IJ}}{\kappa_{pq}} = \pdv{L_{IJ}}{\theta_n} = \pdv{L_{IJ}}{\varphi} = 0.
\end{eqnarray}
From Eq.~\eqref{eq:rotated_state},
we have
\begin{align}\label{eq:identity1}
    \dfrac{\partial \Delta E}{\partial \varphi} = -4 \mathcal{H}_{IJ}, \quad \text{ and} \quad
    \dfrac{\partial \mathcal{H}_{IJ}}{\partial \varphi} = \Delta E,
\end{align}
where $\mathcal{H}_{IJ}  =  \mel{\Psi_I}{ \hat{\mathcal{H}}  }{\Psi_J}$ (we assume that $\hat{\mathcal{H}}$ is a real operator).
We also introduce the two non-zero derivatives of the overlap $\overline{S}_{IJ}$ that read
\begin{equation}
    \begin{split}
         \bra{\Psi_I}\ket{\frac{\partial}{\partial \theta_n} \Psi_J} \neq 0, \quad \text{ and} \quad  \bra{\Psi_I}\ket{\frac{\partial}{\partial \varphi} \Psi_J} = -1.
    \end{split}
\end{equation}
From these simple relations, one obtains the $\overline{\varphi}^{IJ}$
multiplier as
\begin{align}
   \pdv{L_{IJ}}{\varphi} = 0 &= - 1 -4 \overline{\varphi}^{IJ} \longrightarrow  \overline{\varphi}^{IJ} = -\frac{1}{4}.
\end{align}
Combining this result with Eq.~\eqref{eq:NAC_stationary} and Eq.~\eqref{eq:identity1} provides the other stationary equations for the orbital parameters,
\begin{align}
\begin{split}
    \sum_{rs} \overline{\kappa}_{rs}^{IJ} H^{\rm OO}_{pq,rs} + \sum_n\overline{\theta}_n^{IJ} H^{\rm OC}_{pq,n} +  \bra{\Psi_I}\frac{\partial \hat{\mathcal{H}}}{\partial \kappa_{pq}}\ket{\Psi_J} = 0,
\end{split}
\end{align}
and for the ansatze parameters,
\begin{align}
\begin{split}
    & \Delta E \bra{\Psi_I}\ket{\frac{\partial}{\partial \theta_n} \Psi_J}+ \sum_{pq} \overline{\kappa}_{pq}^{IJ} H^{\rm CO}_{n,pq} \\
    & + \sum_m \overline{\theta}_m^{IJ} H^{\rm CC}_{n,m} + \frac{\partial \mathcal{H}_{IJ}}{\partial \theta_n} = 0,
\end{split}
\end{align}
where we multiplied both sides by $\Delta E$. Let us define the orbital and circuit gradients, respectively, as follows:
\begin{align}\label{eq:GOGC}
    {G}^{\text{O},IJ}_{pq} &:=  \bra{\Psi_I}\frac{\partial \hat{\mathcal{H}}}{\partial \kappa_{pq}}\ket{\Psi_J}\\
    {G}^{\text{C},IJ}_{n} &:= \Delta E \bra{\Psi_I}\ket{\frac{\partial}{\partial \theta_n} \Psi_J} +  \frac{\partial \mathcal{H}_{IJ}}{\partial \theta_n} = 0.
\end{align}
Note that the circuit gradient
can actually be set to 0.
Indeed, we have
\begin{align}\label{eq:shittytermiszero}
    \frac{\partial \mathcal{H}_{IJ}}{\partial \theta_n} &=  \mel{\frac{\partial}{\partial \theta_n}\Psi_I}{\hat{\mathcal{H}} }{\Psi_J} + 0 + \mel{\Psi_I}{\hat{\mathcal{H}} }{\frac{\partial}{\partial \theta_n} \Psi_J}\nonumber\\
    &= -\Delta E \bra{\Psi_I}\ket{\frac{\partial}{\partial \theta_n} \Psi_J},
\end{align}
such that the coupled-perturbed equations read
\begin{eqnarray} 
\begin{pmatrix}
\bfH^{\rm OO} & \bfH^{\rm OC} \\
\bfH^{\rm CO} & \bfH^{\rm CC}
\end{pmatrix}
\begin{pmatrix}
\overline{\bmkappa}\\
\overline{\bmtheta} \\
\end{pmatrix}
=
-\begin{pmatrix}
{\bfG}^{\text{O},IJ}\\
0
\end{pmatrix}.
\end{eqnarray}
The final form of the CI term of the NAC can now be written as follows,
\begin{align}
    D_{IJ}^{\rm CI} = \frac{\partial L_{IJ}}{\partial x} = (\Delta E)^{-1} \Bigg(\bra{\Psi_I}\frac{\partial \hat{\mathcal{H}}}{\partial x} \ket{\Psi_J} + \nonumber\\
    \sum_{p,q} \overline{\kappa}_{pq}^{IJ} \dfrac{\partial^2 E_\text{SA}}{\partial \kappa_{pq} \partial x}
    +  \sum_{n} \overline{\theta}_{n}^{IJ} \dfrac{\partial^2 E_\text{SA}}{\partial \theta_n\partial x}\Bigg),
\end{align}
where the last term has been set to zero, similarly as in Eq.~\eqref{eq:shittytermiszero}:
\begin{align}
\overline{\varphi}\dfrac{\partial^2 \Delta E}{\partial \varphi \partial x} &= \frac{\partial}{\partial x} \langle \Psi_I | \hat{\mathcal{H}} | \Psi_J \rangle\nonumber \\
&= -\Delta E \bra{\Psi_I}\ket{\frac{\partial \Psi_J}{\partial x}} + \Delta E \bra{\Psi_I}\ket{\frac{\partial \Psi_J}{\partial x}} = 0.
\end{align}
Similar to the analytical gradient calculation, we end up with:
\begin{equation}\label{eq:appendix_NAC_final}
\begin{split}
    D_{IJ} = \frac{1}{ E_J - E_I } \Bigg( & \sum_{pq} \pdv{h_{pq}}{x} \gamma_{pq}^{IJ,\rm eff} + \dfrac{1}{2}\sum_{pqrs} \pdv{ g_{pqrs}}{x} \Gamma_{pqrs}^{IJ,\rm eff}  \\ 
 + &\sum_K  \sum_n   w_K \overline{\theta}_n^{IJ}   G^{\text{C},K}_n(\tfrac{\partial \hat{\mathcal{H}}}{\partial x}) \Bigg)\\
 - &\dfrac{1}{2}\sum_{pq}\gamma_{pq}^{IJ} \big( (\partial_x p|q) - (q | \partial_x p) \big),
\end{split}
\end{equation}
where the effective transition 1- and 2-RDMs read
\begin{eqnarray}
{\bm \gamma}^{IJ,\text{eff}} &=& {\bm \gamma}^{IJ} + \tilde{\bm \gamma}^{IJ, \rm SA }\\
{\bm \Gamma}^{IJ,\text{eff}} &=& {\bm \Gamma}^{IJ} + \tilde{\bm \Gamma}^{IJ, \rm SA }, 
\end{eqnarray} 
where ${ \gamma}^{IJ}_{pq} = \bra{\Psi_I} \hat{E}_{pq} \ket{\Psi_J} $ and ${ \Gamma}^{IJ}_{pqrs} = \bra{\Psi_I} \hat{e}_{pqrs} \ket{\Psi_J} $ are the transition 1- and 2-RDMs,
and
\begin{eqnarray}
\tilde{\gamma}_{pq}^{IJ, \rm SA } =   \sum_o &\big( \gamma_{oq}^\text{SA} \overline{\kappa}_{op}^{IJ} + \gamma_{po}^{\rm SA} \overline{\kappa}_{oq}^{IJ} \big) \\
\tilde{\Gamma}_{pqrs}^{IJ, \rm SA } = \sum_o & \big( \Gamma^{\rm SA}_{oqrs} \overline{\kappa}_{op}^{IJ} + \Gamma_{pors}^{\rm SA}\overline{\kappa}_{oq}^{IJ} \nonumber \\
& + \Gamma^{\rm SA}_{pqos} \overline{\kappa}_{or}^{IJ} + \Gamma_{pqro}^{\rm SA} \overline{\kappa}_{os}^{IJ} \big)
\end{eqnarray}
are the state-averaged 1- and 2-RDMs  (encoding orbital contributions).

\end{document}